\documentclass[english]{article}
\pdfoutput=1
\usepackage{lmodern}

\usepackage[T1]{fontenc}
\usepackage[latin9]{inputenc}
\usepackage{geometry}
\usepackage{color}
\usepackage{amsmath}
\usepackage{amssymb}
\usepackage{graphicx}

\makeatletter

\newenvironment{lyxlist}[1]
	{\begin{list}{}
		{\settowidth{\labelwidth}{#1}
		 \setlength{\leftmargin}{\labelwidth}
		 \addtolength{\leftmargin}{\labelsep}
		 }}
	{\end{list}}

\@ifundefined{date}{}{\date{}}
\makeatother

\usepackage{babel}
\begin{document}
\title{Symmetry resolved entanglement: Exact results in 1D and beyond}
\author{Shachar Fraenkel and Moshe Goldstein\\
\emph{\normalsize{}Raymond and Beverly Sackler School of Physics and
Astronomy, Tel-Aviv University, Tel Aviv 6997801, Israel}\\
{\normalsize{}E-mail: shacharf@mail.tau.ac.il}}
\maketitle
\begin{abstract}
In a quantum many-body system that possesses an additive conserved
quantity, the entanglement entropy of a subsystem can be resolved
into a sum of contributions from different sectors of the subsystem's
reduced density matrix, each sector corresponding to a possible value
of the conserved quantity. Recent studies have discussed the basic
properties of these symmetry-resolved contributions, and calculated
them using conformal field theory and numerical methods. In this work
we employ the generalized Fisher-Hartwig conjecture to obtain exact
results for the characteristic function of the symmetry-resolved entanglement
(``flux-resolved entanglement'') for certain 1D spin chains, or,
equivalently, the 1D fermionic tight binding and the Kitaev chain
models. These results are true up to corrections of order $o\left(L^{-1}\right)$
where $L$ is the subsystem size. We confirm that this calculation
is in good agreement with numerical results. For the gapless tight
binding chain we report an intriguing periodic structure of the characteristic
functions, which nicely extends the structure predicted by conformal
field theory. For the Kitaev chain in the topological phase we demonstrate
the degeneracy between the even and odd fermion parity sectors of
the entanglement spectrum due to virtual Majoranas at the entanglement
cut. We also employ the Widom conjecture to obtain the leading behavior
of the symmetry-resolved entanglement entropy in higher dimensions
for an ungapped free Fermi gas in its ground state.
\end{abstract}


\section{Introduction}

The importance of entanglement to the analysis of quantum systems
can hardly be exaggerated. In the context of many-body systems, the
study of entanglement can help to identify important phenomena such
as quantum phase transitions \cite{Osterloh_2002,PhysRevLett.90.227902,RevModPhys.80.517,2009JPhA...42X4005C,LAFLORENCIE20161},
to point out systems that can provide efficient resources for quantum
information processing \cite{PhysRevLett.70.1895,doi:10.1137/S0097539795293172,PhysRevLett.80.2245,RevModPhys.74.145,PhysRevA.69.052308,PhysRevLett.103.150502},
and to determine the applicability of methods that are based on tensor
networks \cite{RevModPhys.77.259,doi:10.1080/14789940801912366}.

The main quantitative measure of entanglement in a many-body system
is the entanglement entropy (EE) \cite{LAFLORENCIE20161}. For a many-body
system in a pure state $|\psi\rangle$, we define the density matrix
of the system as
\begin{equation}
\rho=|\psi\rangle\langle\psi|.
\end{equation}
Let $A$ be a subsystem, while the rest of the system will be denoted
by $B$. The reduced density matrix (RDM) of subsystem $A$ will then
be defined as
\begin{equation}
\rho_{A}=\text{Tr}_{B}\left(\rho\right),
\end{equation}
where $\text{Tr}_{B}$ is the partial trace over the degrees of freedom
of subsystem $B$. We define the $n$th moment of the reduced density
matrix of $A$, which we will subsequently refer to as the $n$th
R\'enyi entanglement entropy (REE), as
\begin{equation}
S_{n}=\text{Tr}\left(\rho_{A}^{n}\right).\label{eq: Renyi definition}
\end{equation}
Note that this definition of the REE is different than the usual one,
$S_{n}=\frac{1}{1-n}\log\left(\text{Tr}\left(\rho_{A}^{n}\right)\right)$.
We further define the von-Neumann entanglement entropy (vNEE) of $A$
\cite{PPN379400774} as
\begin{equation}
\mathcal{S}=-\text{Tr}\left(\rho_{A}\log\rho_{A}\right)=-\lim_{n\rightarrow1}\partial_{n}S_{n}.\label{eq: vNEE definition}
\end{equation}
The quantities defined in (\ref{eq: Renyi definition}) and (\ref{eq: vNEE definition})
are the two fundamental types of EE, and they constitute important
tools for understanding entanglement, in particular in the field of
quantum information \cite{PhysRevA.53.2046,bennett2000quantum,Latorre:2004:GSE:2011572.2011576,RevModPhys.81.865}.

We consider the case where the entire system is characterized by some
fixed value of a conserved charge $\hat{Q}$, so that the density
matrix $\rho$ commutes with $\hat{Q}$. We assume that the total
charge $\hat{Q}$ can be written as $\hat{Q}=\hat{Q}_{A}+\hat{Q}_{B}$,
where $\hat{Q}_{i}$ is the contribution of subsystem $i$ to the
total charge. Applying the partial trace over subsystem $B$ to the
equation $\left[\hat{Q},\rho\right]=0$, we obtain
\begin{equation}
\left[\hat{Q}_{A},\rho_{A}\right]=0,
\end{equation}
which means that $\rho_{A}$ is block-diagonal with respect to the
eigenbasis of $\hat{Q}_{A}$. In such a representation, each block
(charge sector) corresponds to an eigenvalue $Q_{A}$ of $\hat{Q}_{A}$,
and we can therefore denote this block by $\rho_{A}^{\left(Q_{A}\right)}$,
and define for each such eigenvalue \cite{Laflorencie_2014,PhysRevLett.120.200602,PhysRevB.98.041106,PhysRevLett.121.150501,PhysRevA.100.022324}
\begin{align}
S_{n}\left(Q_{A}\right) & =\text{Tr}\left(\left(\rho_{A}^{\left(Q_{A}\right)}\right)^{n}\right),\nonumber \\
\mathcal{S}\left(Q_{A}\right) & =-\text{Tr}\left(\rho_{A}^{\left(Q_{A}\right)}\log\rho_{A}^{\left(Q_{A}\right)}\right)=-\lim_{n\rightarrow1}\partial_{n}S_{n}\left(Q_{A}\right),\label{eq: charge-resolved definition}
\end{align}
which are named the symmetry-resolved REE and the symmetry-resolved
vNEE, respectively. It is evident that these quantities satisfy $\mathcal{S}=\sum_{Q_{A}}\mathcal{S}\left(Q_{A}\right)$
and $S_{n}=\sum_{Q_{A}}S_{n}\left(Q_{A}\right)$. Note that some works
normalize each block by each trace \cite{PhysRevB.98.041106,PhysRevLett.121.150501,PhysRevA.100.022324}
before calculating the entropies, which thus quantify the entanglement
after a projective charge measurement. We prefer not to do so and
instead use (\ref{eq: charge-resolved definition}), following \cite{Laflorencie_2014,PhysRevLett.120.200602},
because the resulting resolved entropies, while not
entanglement measures by themselves, are not only more accessible
to calculations, but are also directly experimentally measurable,
using either the replica trick \cite{PhysRevLett.120.200602,PhysRevA.98.032302,PhysRevA.99.062309},
or random time evolution which conserves the charge \cite{PhysRevLett.120.050406,PhysRevA.97.023604}.
Let us note that $S_{1}\left(Q_{A}\right)$ is simply the distribution
$P\left(Q_{A}\right)$ of charge in subsystem $A$. Using
this, one may easily employ our results to find the normalized versions
of the REE and vNEE, whose roles and limitations as entanglement measures
are discussed in \cite{PhysRevA.100.022324}.

When $\hat{Q}$ can assume any integer value (e.g., when particle
number or total $S_{z}$ are conserved), we define the flux-resolved
REE as
\begin{equation}
S_{n}\left(\alpha\right)=\text{Tr}\left(\rho_{A}^{n}e^{i\alpha\hat{Q}_{A}}\right).\label{eq: flux resolved REE}
\end{equation}
The importance of this quantity arises from the fact that it is the
characteristic function related to the symmetry-resolved REE via Fourier
transform \cite{PhysRevLett.120.200602}:
\begin{equation}
S_{n}\left(Q_{A}\right)=\underset{-\pi}{\overset{\pi}{\int}}\frac{d\alpha}{2\pi}S_{n}\left(\alpha\right)e^{-i\alpha Q_{A}}.\label{eq: flux-resolved to charge-resolved}
\end{equation}
The flux-resolved and charge-resolved REEs have previously been approximately
calculated for 1D many-body systems using conformal field theory (CFT)
and numerical techniques \cite{Laflorencie_2014,PhysRevLett.120.200602,PhysRevB.98.041106,PhysRevLett.121.150501,PhysRevA.100.022324}.

The flux-resolved REE has an analog for discrete symmetries, i.e.,
when the quantity conserved is $Q\mod p$ where $p$ is some natural
number (e.g., fermion parity for $p=2$) \cite{PhysRevLett.120.200602}.
In this case we define 
\begin{equation}
S_{n}\left(\alpha\right)=\text{Tr}\left(\rho_{A}^{n}e^{i\frac{2\pi\alpha}{p}\hat{Q}_{A}}\right),\,\,\alpha=0,1,\ldots,p-1,\label{eq: discrete flux resolved REE}
\end{equation}
and then
\begin{equation}
S_{n}\left(Q_{A}\right)=\frac{1}{p}\underset{\alpha=0}{\overset{p-1}{\sum}}e^{-i\frac{2\pi\alpha}{p}Q_{A}}S_{n}\left(\alpha\right),\,\,Q_{A}=0,1,\ldots,p-1.
\end{equation}

The study of the symmetry-resolved entanglement also sheds light on
the attributes of the entanglement spectrum. The latter is the spectrum
of the entanglement Hamiltonian $H_{A}$ of subsystem $A$, defined
through $\rho_{A}=\exp\left(-H_{A}\right)$. It is especially interesting
in topological systems, which are often characterized by a bulk gap
and topologically-protected gapless edge excitations \cite{Wen:2004ym}.
The entanglement Hamiltonian generically possesses \textquotedblleft low
energy\textquotedblright{} modes at its virtual edge (the boundary
between the subsystem and the rest of the system) similar to those
the physical Hamiltonian possesses at a physical edge \cite{LAFLORENCIE20161,PhysRevLett.101.010504}.
In particular, starting with the seminal work of Kitaev \cite{Kitaev_2001},
a lot of theoretical and experimental effort is currently directed
at realizing systems with topologically-protected Majorana zero-modes
in 1D \cite{PhysRevLett.105.077001,PhysRevLett.105.177002} or above
\cite{PhysRevLett.100.096407,PhysRevLett.101.120403}, which could
serve as a resource for topological quantum computation \cite{sarma2015majorana}.
Similar Majorana zero-modes should show up in the entanglement spectrum
\cite{PhysRevB.83.075102,PhysRevB.99.115429,universe5010033}.

This work presents a calculation of the asymptotic behavior of the
flux-resolved and the symmetry-resolved EE for a (large) subsystem
of an infinite 1D spin chain in its ground state, or of equivalent
fermionic chains, as well as the leading order behavior for free fermions
in higher dimensions, using the generalized Fisher-Hartwig \cite{10.2307/23030524}
and Widom \cite{widom1982class} conjectures, respectively. Section
\ref{sec:Model-and-main} presents the 1D model and summarizes the
main results pertaining to it. Section \ref{sec:Asymptotics-of-the}
is a summary of previously obtained results for the non-resolved entanglement,
upon which our calculations will rely. In Section \ref{sec:Symmetry-resolved-entanglement-e}
we discuss the asymptotics of the flux-resolved REE in a 1D spin chain
with rotational symmetry in the plane perpendicular to the magnetic
field, or in a gapless tight-binding chain with conserved fermion
number, and show that the result has a periodic structure that is
a natural extension of the CFT results. In Section \ref{sec:Symmetry-resolved-entanglement XY}
we derive analytical results for the symmetry-resolved REE and vNEE
in the case where the system has no such rotational symmetry, but
the parity of the number of up spins is still maintained. This maps
into the fermionic Kitaev chain, where fermion number is not conserved
but parity is. We find that the fermion parity even and odd entanglement
spectra become degenerate due to the appearance of Majorana entanglement
zero-modes in the topological phase, but not in the trivial phase.
At the critical point separating these phases a power
law arises, in agreement with CFT results. Section \ref{sec:Generalization-to-higher}
addresses the leading behavior of the charge-resolved REE in an ungapped
free Fermi gas in a general dimension. Finally, Section \ref{sec:Conclusions-and-future}
presents our conclusions and an outlook for the future.

\section{\label{sec:Model-and-main}Model and main results for 1D}

The 1D model discussed in this work is that of a spin chain in a transverse
magnetic field. This system is described by the Hamiltonian
\begin{equation}
\mathcal{H}=-J\underset{m=-N/2}{\overset{N/2-1}{\sum}}\left[\left(1+\gamma\right)\sigma_{m}^{x}\sigma_{m+1}^{x}+\left(1-\gamma\right)\sigma_{m}^{y}\sigma_{m+1}^{y}\right]-Jh\underset{m=-N/2}{\overset{N/2}{\sum}}\sigma_{m}^{z},\label{eq: General Hamiltonian}
\end{equation}
where $\sigma_{m}^{x}$, $\sigma_{m}^{y}$ and $\sigma_{m}^{z}$
are Pauli matrices for a spin-$\frac{1}{2}$ at lattice site{}
$m=-\frac{N}{2},\ldots,\frac{N}{2}$, $N+1$ being the total number
of sites ($N$ is assumed to be even), $J$
is the exchange interaction scale, $h$ is the dimensionless magnetic
field, and $\gamma$ is the dimensionless anisotropy
parameter. Without loss of generality we may assume
$J>0$ and $\gamma\ge0$. For $\gamma=0$ the system is isotropic,
i.e., has rotational symmetry in the XY plane; the isotropic case
is called the XX model, while the general case $\gamma\neq0$ is named
the XY model. We focus on an infinite chain ($N\rightarrow\infty$),
and on asymptotic results that are valid for a subsystem of $L$ contiguous
sites where $L\gg1$.

The treatment of the system relies on the Jordan-Wigner transformation
of ${\cal H}$ \cite{LIEB1961407}. We introduce two Majorana operators
for each site on the spin chain:
\begin{equation}
c_{2l-1}=\left(\underset{n=-N/2}{\overset{l-1}{\Pi}}\sigma_{n}^{z}\right)\sigma_{l}^{x}\,\,\text{and}\,\,c_{2l}=\left(\underset{n=-N/2}{\overset{l-1}{\Pi}}\sigma_{n}^{z}\right)\sigma_{l}^{y}.
\end{equation}
We then define for each $-N/2\le m\le N/2$
\begin{equation}
a_{m}=\frac{1}{2}\left(c_{2m-1}-ic_{2m}\right).\label{eq: JW fermions}
\end{equation}
The operators $a_{m}$ obey fermionic anti-commutation relations (i.e.,
$\left\{ a_{m},a_{n}^{\dagger}\right\} =\delta_{mn}$ and $\left\{ a_{m},a_{n}\right\} =0$),
and in the terms of these operators $\mathcal{H}$ is written as
\begin{equation}
\mathcal{H}=2J\underset{m=-N/2}{\overset{N/2-1}{\sum}}\left[a_{m}^{\dagger}a_{m+1}+a_{m+1}^{\dagger}a_{m}+\gamma\left(a_{m}^{\dagger}a_{m+1}^{\dagger}+a_{m+1}a_{m}\right)\right]-2Jh\underset{m=-N/2}{\overset{N/2}{\sum}}\left(a_{m}^{\dagger}a_{m}-\frac{1}{2}\right).\label{eq: JW Hamiltonian}
\end{equation}
Now the Hamiltonian is described in terms of a quadratic chain of
spinless fermions, the Kitaev chain \cite{Kitaev_2001}. The system
can be solved exactly using a Fourier transform of $a_{m}$ followed
by a Bogoliubov transformation. This allows us to show that the system
has a unique\footnote{The ground state is unique (up to edge effects, which are discussed below) as long as $h\neq2\sqrt{1-\gamma^{2}}$;
for $h=2\sqrt{1-\gamma^{2}}$ it is doubly degenerate \cite{Franchini_2007}.} ground state $|GS\rangle$, and also to obtain its spectrum at
the limit $N\rightarrow\infty$ \cite{Franchini_2007}:
\begin{equation}
\varepsilon_{\theta}=4J\sqrt{\left(\cos \theta-\frac{h}{2}\right)^{2}+\gamma^{2}\sin^{2}\theta}\,,\,\,\,0\le \theta\le 2\pi.
\end{equation}
We assume that the system is at its ground state, i.e., $\rho=|GS\rangle\langle GS|$.

In the case of the XX model, the system satisfies the conservation
of the total fermionic number (total spin in the $z$ direction):{}
$Q=\underset{m=-N/2}{\overset{N/2}{\sum}}a_{m}^{\dagger}a_{m}=\underset{m=-N/2}{\overset{N/2}{\sum}}\frac{1}{2}\left(\sigma_{m}^{z}+1\right)$.
We can therefore define $S_{n}\left(\alpha\right)$ for a subsystem
of $L$ sites using the definition for non-discrete symmetries in
(\ref{eq: flux resolved REE}). In this case, for $\left|h\right|\le2$,
the system is also gapless with the Fermi points being at $\pm k_{F}$,
where
\begin{equation}
k_{F}\equiv\arccos\left(\frac{h}{2}\right).
\end{equation}

In the case of the XY model, however, $Q$ is no longer a conserved
quantity of the system. Nevertheless, the system is still characterized
by a discrete symmetry: since the total fermionic number can only
change by even numbers, its parity $\left(-1\right)^{Q}$ is in fact
conserved. Thus the RDM of subsystem $A$ can be decomposed into two
sectors, corresponding to odd and even values of $Q_{A}$. Following
the definition of the analog of the flux-resolved REE for discrete
symmetries in (\ref{eq: discrete flux resolved REE}), we define
\begin{equation}
S_{n}^{\left(-\right)}\equiv\text{Tr}\left(\rho_{A}^{n}\left(-1\right)^{\hat{Q}_{A}}\right),
\end{equation}
and decompose the REE by writing
\begin{equation}
S_{n}=S_{n}^{\left(\mathrm{even}\right)}+S_{n}^{\left(\mathrm{odd}\right)},
\end{equation}
where 
\begin{equation}
S_{n}^{\left(\mathrm{even}\right)}\equiv\frac{1}{2}\left[S_{n}+S_{n}^{\left(-\right)}\right]\,\,\text{and}\,\,S_{n}^{\left(\mathrm{odd}\right)}\equiv\frac{1}{2}\left[S_{n}-S_{n}^{\left(-\right)}\right],
\end{equation}
with similar definitions for the vNEEs $\mathcal{S}^{\left(-\right)}$,
$\mathcal{S}^{\left(\mathrm{even}\right)}$ and $\mathcal{S}^{\left(\mathrm{odd}\right)}$.

For the XY model, the system is gapped for $\left|h\right|\neq2$,
while at $h=\pm2$ the gap closes and a phase transition occurs. For
$\left|h\right|<2$ the system is in a topologically nontrivial phase
with Majorana edge-modes at its real edges, while for $\left|h\right|>2$
it is found in a topologically trivial phase with no Majorana edge-modes
\cite{Kitaev_2001}. 

\subsection{Results for the XX model}

Assuming $\left|h\right|\le2$, we write $\mathcal{L}\equiv2L\left|\sin k_{F}\right|$
and define a natural number $m_{c}=m_{c}\left(n\right)\equiv\lceil\frac{n}{4}\rceil+1$.
We will show that for $\mathcal{L}\gg1$, 
\begin{equation}
S_{n}\left(\alpha\right)=\exp\left[i\frac{k_{F}}{\pi}\alpha L+\left[\frac{1}{6}\left(\frac{1}{n}-n\right)-\frac{\alpha^{2}}{2\pi^{2}n}\right]\ln\mathcal{L}+\Upsilon_{0}\left(n,\alpha\right)+\Upsilon_{1}\left(n,\alpha,L,k_{F}\right)+o\left(\mathcal{L}^{-1}\right)\right],
\end{equation}
where
\begin{equation}
\Upsilon_{0}\left(n,\alpha\right)\equiv-\frac{1}{\pi^{2}}\underset{0}{\overset{\infty}{\int}}\ln\left[\frac{2\cos\alpha+2\cosh\left(nu\right)}{\left(2\cosh\left(\frac{u}{2}\right)\right)^{2n}}\right]du\underset{0}{\overset{\infty}{\int}}\left[\frac{e^{-t}}{t}-\frac{\cos\left(\frac{ut}{2\pi}\right)}{2\sinh\left(\frac{t}{2}\right)}\right]dt,
\end{equation}
and
\begin{align}
\Upsilon_{1}\left(n,\alpha,L,k_{F}\right) & \equiv\underset{m=1}{\overset{m_{c}}{\sum}}\ln\left[1+\mathcal{L}^{-\frac{2}{n}\left(2m-1-\frac{\alpha}{\pi}\right)}e^{-2ik_{F}L}\frac{\Gamma\left(\frac{1}{2}+\frac{1}{2n}\left(2m-1-\frac{\alpha}{\pi}\right)\right)^{2}}{\Gamma\left(\frac{1}{2}-\frac{1}{2n}\left(2m-1-\frac{\alpha}{\pi}\right)\right)^{2}}\right]+\nonumber \\
 & +\underset{m=1}{\overset{m_{c}}{\sum}}\ln\left[1+\mathcal{L}^{-\frac{2}{n}\left(2m-1+\frac{\alpha}{\pi}\right)}e^{2ik_{F}L}\frac{\Gamma\left(\frac{1}{2}+\frac{1}{2n}\left(2m-1+\frac{\alpha}{\pi}\right)\right)^{2}}{\Gamma\left(\frac{1}{2}-\frac{1}{2n}\left(2m-1+\frac{\alpha}{\pi}\right)\right)^{2}}\right].
\end{align}
The term $\left[\frac{1}{6}\left(\frac{1}{n}-n\right)-\frac{\alpha^{2}}{2\pi^{2}n}\right]\ln\mathcal{L}$
in the exponent has been already found before, using CFT techniques
\cite{PhysRevLett.120.200602,PhysRevB.98.041106}, and our calculation
not only derives it rigorously, but also completes the picture up
to corrections of order $o\left(\mathcal{L}^{-1}\right)$.

Furthermore, in \ref{subsec: Periodic structure} we will see that
this result can be written as
\begin{equation}
S_{n}\left(\alpha\right)=\underset{j=-m_{c}}{\overset{m_{c}}{\sum}}\tilde{S}_{n}\left(\alpha+2\pi j\right)+o\left(\mathcal{L}^{-1}\right),
\end{equation}
where $\tilde{S}_{n}$ is an analytic function that is defined on
the entire real line. This shows that $S_{n}\left(\alpha\right)$
has a structure that is natural in the context of CFT, as we explain
below.

\subsection{Results for the XY model}

We will use the notations $k\equiv\gamma\slash\sqrt{\left(h/2\right)^{2}+\gamma^{2}-1}$
and $k'\equiv\sqrt{1-k^{2}}$, and denote by $k_{n}$ the positive
solution to the equation $q^{n}=\exp\left[-\pi I\left(\sqrt{1-k_{n}^{2}}\right)\slash I\left(k_{n}\right)\right]$,
where
\begin{equation}
I\left(k\right)=\underset{0}{\overset{1}{\int}}\frac{dx}{\sqrt{\left(1-x^{2}\right)\left(1-k^{2}x^{2}\right)}}
\end{equation}
is the complete elliptic integral of the first kind and $q\equiv\exp\left[-\pi I\left(k'\right)\slash I\left(k\right)\right]$
is the nome \cite{NIST:DLMF}. Assuming that $0\le h\neq2$, we will
find that as $L\rightarrow\infty$,
\begin{equation}
\underset{L\rightarrow\infty}{\lim}\left(-1\right)^{L}S_{n}^{\left(-\right)}=\begin{cases}
0, & h<2\\
\left[\frac{\left(k\cdot k'\right)^{2n}\left(1-k_{n}^{2}\right)^{2}}{16^{n-1}k_{n}^{2}}\right]^{\frac{1}{12}}, & h>2
\end{cases},\label{eq: XY REE summary}
\end{equation}
and

\begin{equation}
\underset{L\rightarrow\infty}{\lim}\left(-1\right)^{L}\mathcal{S}^{\left(-\right)}=\begin{cases}
0, & h<2\\
\frac{\sqrt{k'}}{3}\left[\ln2-\frac{1}{2}\ln\left(k\cdot k'\right)-\frac{I\left(k\right)I\left(k'\right)}{\pi}\left(1+k^{2}\right)\right], & h>2
\end{cases}.\label{eq: XY vNEE summary}
\end{equation}
For finite $L$, the corrections to these expressions are exponentially
small in $L$. We are not aware of extensions of the
Fisher-Hartwig conjecture which allow to calculate these corrections,
but we verify numerically that they are negligible even for relatively
small values of $L$. For $h<2$ we get in particular that $\underset{L\rightarrow\infty}{\lim}\left[\mathcal{S}^{\left(\mathrm{even}\right)}-\mathcal{S}^{\left(\mathrm{odd}\right)}\right]=0$,
due to a degeneracy between the spectra of the even charge sector
and the odd charge sector. This degeneracy stems from the appearance
of Majorana zero-modes at the virtual edges of the entanglement Hamiltonian.

For the critical field $h=2$, $S_{n}^{\left(-\right)}$
and $\mathcal{S}^{\left(-\right)}$ still vanish as $L\rightarrow\infty$,
but only as a power law, rather than exponentially. We will find that
in this case there is a positive factor $A\left(n,\gamma\right)$
such that we can write the following leading order approximation for
large $L$:
\begin{equation}
\left(-1\right)^{L}S_{n}^{\left(-\right)}\approx A\left(n,\gamma\right)L^{-\frac{1}{6n}-\frac{n}{12}},
\end{equation}
and
\begin{equation}
\left(-1\right)^{L}\mathcal{S}^{\left(-\right)}\approx-\frac{A\left(1,\gamma\right)}{12}L^{-\frac{1}{4}}\ln L,\label{eq: Gapless XY vNEE summary}
\end{equation}
in accordance with the CFT results of \cite{PhysRevLett.120.200602}.

These results can be extended to $h<0$ by plugging
in the corresponding result for $\left|h\right|$, only in this case
the $\left(-1\right)^{L}$ factor that appears in{}
(\ref{eq: XY REE summary})--(\ref{eq: Gapless XY vNEE summary})
is absent.

\section{\label{sec:Asymptotics-of-the}Asymptotics of the spectrum of the
RDM in 1D}

For the convenience of the reader, this section summarizes results
from previous works that will be instrumental to the calculations
that follow, and were originally presented in \cite{Franchini_2007,2004JSP...116...79J,Its_2005,Calabrese_2010,jin2007entropy}.

\subsection{The subsystem correlation matrix}

The Jordan-Wigner transformation constitutes the basis for the calculation
of the EE for a subsystem $A$ of $L$ sites \cite{2004JSP...116...79J,Its_2005}.
One can show that the Majorana operators $c_{n}$ that belong to subsystem
$A$ obey
\begin{equation}
\langle GS|c_{n}|GS\rangle=0,\,\,\langle GS|c_{m}c_{n}|GS\rangle=\delta_{mn}+i\left(B_{L}\right)_{mn};\,\,m,n=1,\ldots,2L.
\end{equation}
Here $B_{L}$ is a $2L\times2L$ matrix defined as
\begin{equation}
B_{L}=\left(\begin{array}{cccc}
\Pi_{0} & \Pi_{-1} & \cdots & \Pi_{1-L}\\
\Pi_{1} & \Pi_{0} &  & \vdots\\
\vdots &  & \ddots & \vdots\\
\Pi_{L-1} & \cdots & \cdots & \Pi_{0}
\end{array}\right),\,\,\,\Pi_{m}\equiv\frac{1}{2\pi}\underset{0}{\overset{2\pi}{\int}}d\theta e^{-im\theta}\mathcal{G}\left(\theta\right),\label{eq: subsystem correlation matrix}
\end{equation}
where 
\begin{equation}
\mathcal{G}\left(\theta\right)\equiv\left(\begin{array}{cc}
0 & g\left(\theta\right)\\
-g^{-1}\left(\theta\right) & 0
\end{array}\right),\,\,\,g\left(\theta\right)\equiv\frac{\cos\theta-i\gamma\sin\theta-\frac{h}{2}}{\left|\cos\theta-i\gamma\sin\theta-\frac{h}{2}\right|}.\label{eq: Toeplitz generating function}
\end{equation}
Using an orthogonal matrix $V$ we can transform $B_{L}$ into the
form 
\begin{equation}
VB_{L}V^{T}=\oplus_{m=1}^{L}\nu_{m}\left(\begin{array}{cc}
0 & 1\\
-1 & 0
\end{array}\right),
\end{equation}
where $\nu_{m}$ are real numbers which satisfy $-1<\nu_{m}<1$. We
use $V$ to transform the Majorana operators as well by defining
\begin{equation}
d_{m}=\underset{n=1}{\overset{2L}{\sum}}V_{mn}c_{n},\,\,m=1,\ldots,2L.
\end{equation}
Similarly to the transformation of $c_{n}$ into fermionic operators
in (\ref{eq: JW fermions}), one can obtain a set of $L$ fermionic
operators by introducing $b_{m}\equiv\frac{1}{2}\left(d_{2m}+id_{2m-1}\right)$.
In \cite{2004JSP...116...79J,Peschel_2003} it was shown that the
reduced density matrix of subsystem $A$ in the ground state of the
entire system can be represented by a quite simple expression involving
the fermionic operators $b_{m}$:
\begin{equation}
\rho_{A}=\text{Tr}_{B}\left(|GS\rangle\langle GS|\right)=\underset{m=1}{\overset{L}{\Pi}}\left[\left(\frac{1+\nu_{m}}{2}\right)b_{m}^{\dagger}b_{m}+\left(\frac{1-\nu_{m}}{2}\right)b_{m}b_{m}^{\dagger}\right].\label{eq: RDM using fermionic operators}
\end{equation}

\subsection{Fisher-Hartwig conjecture}

Since the values $\nu_{m}$ in (\ref{eq: RDM using fermionic operators})
determine the spectrum of the RDM $\rho_{A}$, considerable efforts
were invested in estimating them under certain conditions. The general
assumption upon which we will rely is that $L\gg1$. This allows us
to use special cases of the Fisher-Hartwig conjecture \cite{10.2307/23030524}
in order to obtain asymptotic expressions for the EE.

\subsubsection{XX model}

We first consider the isotropic case $\gamma=0$, assuming that $\left|h\right|\le2$
(ungapped chain). In this case, further simplification of the expression
for the correlation matrix $B_{L}$ in (\ref{eq: subsystem correlation matrix})
can be achieved by noticing that
\begin{equation}
B_{L}=G_{L}\otimes\left(\begin{array}{cc}
0 & 1\\
-1 & 0
\end{array}\right),
\end{equation}
with
\begin{equation}
G_{L}=\left(\begin{array}{cccc}
\phi_{0} & \phi_{-1} & \cdots & \phi_{1-L}\\
\phi_{1} & \phi_{0} &  & \vdots\\
\vdots &  & \ddots & \vdots\\
\phi_{L-1} & \cdots & \cdots & \phi_{0}
\end{array}\right),\,\,\,\phi_{m}\equiv\frac{1}{2\pi}\underset{0}{\overset{2\pi}{\int}}d\theta e^{-im\theta}\phi\left(\theta\right),\label{eq: Simplified correlation matrix}
\end{equation}
where we have defined
\begin{equation}
\phi\left(\theta\right)\equiv\begin{cases}
1 & -k_{F}<\theta<k_{F}\\
-1 & k_{F}<\theta<2\pi-k_{F}
\end{cases}\,\text{and}\,k_{F}\equiv\arccos\left(\frac{h}{2}\right).
\end{equation}
The required values $\nu_{m}$ are therefore just the eigenvalues
of the matrix $G_{L}$, or equivalently the zeros of the determinant
$D_{L}\left(\lambda\right)\equiv\det\left(\lambda I_{L}-G_{L}\right)$.
In \cite{2004JSP...116...79J} it was shown that for large $L$, $D_{L}\left(\lambda\right)$
can be written asymptotically as 
\begin{equation}
D_{L}\left(\lambda\right)\sim D_{L}^{\left(0\right)}\left(\lambda\right)\equiv\mathcal{L}^{-2\beta^{2}\left(\lambda\right)}\left[\left(\lambda+1\right)\left(\frac{\lambda+1}{\lambda-1}\right)^{-k_{F}/\pi}\right]^{L}\left[G\left(1+\beta\left(\lambda\right)\right)G\left(1-\beta\left(\lambda\right)\right)\right]^{2}.\label{eq: FH leading}
\end{equation}
Here $\beta\left(\lambda\right)\equiv\frac{1}{2\pi i}\ln\left(\frac{\lambda+1}{\lambda-1}\right)$,
$\mathcal{L}\equiv2L\left|\sin k_{F}\right|$ and $G$ is the Barnes
G-function \cite{NIST:DLMF}. Subleading corrections may be obtained
from the generalized Fisher-Hartwig conjecture \cite{Calabrese_2010}:
\begin{equation}
D_{L}\left(\lambda\right)=\left[\left(\lambda+1\right)\left(\frac{\lambda+1}{\lambda-1}\right)^{-k_{F}/\pi}\right]^{L}\underset{m\in\mathbb{Z}}{\sum}e^{-2imk_{F}L}\mathcal{L}^{-2\left(\beta\left(\lambda\right)+m\right)^{2}}\left[G\left(1+\beta\left(\lambda\right)+m\right)G\left(1-\beta\left(\lambda\right)-m\right)\right]^{2}.\label{eq: FH subleading}
\end{equation}

\subsubsection{XY model\label{subsec:XY-model}}

We now consider the more general case $\gamma\neq0$, i.e., the anisotropic
spin chain. We assume that $h\ge0$ and focus first
on the gapped case $h\neq2$. The system exhibits a quantum phase
transition at $h=2$, and therefore we must separate the cases $h<2$
and $h>2$. We define a number $\sigma$ such that $\sigma=1$ for
$h<2$ and $\sigma=0$ for $h>2$. Following \cite{Franchini_2007},
we also define
\begin{equation}
k\equiv\begin{cases}
\sqrt{\left(1-\left(h/2\right)^{2}-\gamma^{2}\right)\slash\left(1-\left(h/2\right)^{2}\right)}, & h^{2}<4\left(1-\gamma^{2}\right)\\
\sqrt{\left(h/2\right)^{2}+\gamma^{2}-1}\slash\gamma, & 4\left(1-\gamma^{2}\right)<h^{2}<4\\
\gamma\slash\sqrt{\left(h/2\right)^{2}+\gamma^{2}-1}, & h>2
\end{cases},\label{eq: Definition of k}
\end{equation}
and
\begin{equation}
\tau_{0}\equiv I\left(\sqrt{1-k^{2}}\right)\slash I\left(k\right),\label{eq: definition of tau_0}
\end{equation}
where $I\left(k\right)$ is the complete elliptic integral of the
first kind,
\begin{equation}
I\left(k\right)=\underset{0}{\overset{1}{\int}}\frac{dx}{\sqrt{\left(1-x^{2}\right)\left(1-k^{2}x^{2}\right)}}.\label{eq: First elliptic integral}
\end{equation}

In the XY model, a calculation of a different determinant than that
of the XX model is required. Let us define the determinant
\begin{equation}
\tilde{D}_{L}\left(\lambda\right)\equiv\det\left(i\lambda I_{2L}-B_{L}\right)=\underset{m=1}{\overset{L}{\Pi}}\left(\nu_{m}^{2}-\lambda^{2}\right),\label{eq: definition of XY determinant}
\end{equation}
the zeros of which are simply $\pm\nu_{m}$. It was shown in \cite{Its_2005}
that in the large $L$ limit, the following asymptotic expression
for $\tilde{D}_{L}\left(\lambda\right)$ is obtained:
\begin{equation}
\tilde{D}_{L}\left(\lambda\right)\sim\frac{\left(1-\lambda^{2}\right)^{L}}{\varTheta_{3}^{2}\left(\frac{i\sigma\tau_{0}}{2}\right)}\varTheta_{3}\left(\beta\left(\lambda\right)+\frac{i\sigma\tau_{0}}{2}\right)\varTheta_{3}\left(\beta\left(\lambda\right)-\frac{i\sigma\tau_{0}}{2}\right).\label{eq: Asymptotic det XY}
\end{equation}
Here we have defined $\varTheta_{3}\left(s\right)\equiv\vartheta_{3}\left(\pi s,e^{-\pi\tau_{0}}\right)$,
where $\vartheta_{3}\left(z,q\right)=\underset{m=-\infty}{\overset{\infty}{\sum}}q^{m^{2}}e^{2izm}$
is the third Jacobi theta function \cite{whittaker_watson_1996}.
The asymptotic expression for $\tilde{D}_{L}\left(\lambda\right)$
in (\ref{eq: Asymptotic det XY}) has a double zero at each of the
points
\begin{equation}
\lambda_{l}=\tanh\left[\left(l+\frac{1-\sigma}{2}\right)\pi\tau_{0}\right],\,\,l\in\mathbb{Z}.\label{eq: Limit of nu_m}
\end{equation}
This shows that as $L\rightarrow\infty$, the values $\pm\nu_{m}$
are divided into pairs $\tilde{\nu}_{2l-1},\tilde{\nu}_{2l}$ such
that for every $l\in\mathbb{Z}$, $\tilde{\nu}_{2l-1},\tilde{\nu}_{2l}\rightarrow\lambda_{l}$.
Corrections to the asymptotic expression (\ref{eq: Asymptotic det XY})
vanish exponentially as $L\rightarrow\infty$ \cite{jin2007entropy}.

The asymptotics of $\tilde{D}_{L}\left(\lambda\right)$
in the gapless case $h=2$ differs considerably, due to a discontinuity
of the symbol $\mathcal{G}\left(\theta\right)$ that was defined in
(\ref{eq: Toeplitz generating function}). Based on a general conjecture
presented in \cite{PhysRevA.92.042334,PhysRevA.97.062301} and verified there numerically
for several cases, we can predict the two leading terms in the large
$L$ approximation of $\ln\tilde{D}_{L}\left(\lambda\right)$:
\begin{equation}
\ln\tilde{D}_{L}\left(\lambda\right)\sim\ln\left(1-\lambda^{2}\right)L-2\beta^{2}\left(\lambda\right)\ln L.\label{eq: Asymptotic det gapless XY}
\end{equation}
We present the derivation of the above expression in subsection \ref{subsec: Appendix 4}
of the appendix.

\section{\label{sec:Symmetry-resolved-entanglement-e}Symmetry-resolved EE
for the XX model}

Throughout this section we assume that $\gamma=0$ and $\left|h\right|\le2$,
which corresponds to the gapless XX model.

\subsection{Leading order approximation for flux-resolved EE}

From the expression for $\rho_{A}$ in (\ref{eq: RDM using fermionic operators})
we can deduce that the flux-resolved REE may be written as 
\begin{equation}
S_{n}\left(\alpha\right)=\underset{m=1}{\overset{L}{\Pi}}\left[\left(\frac{1+\nu_{m}}{2}\right)^{n}e^{i\alpha}+\left(\frac{1-\nu_{m}}{2}\right)^{n}\right],\label{eq: flux-resolved REE explicit}
\end{equation}
where $\nu_{m}$ are the eigenvalues of the matrix $G_{L}$ defined
in (\ref{eq: Simplified correlation matrix}) \cite{PhysRevLett.120.200602}.

\begin{figure}[h]
\centering{}\includegraphics[scale=0.3]{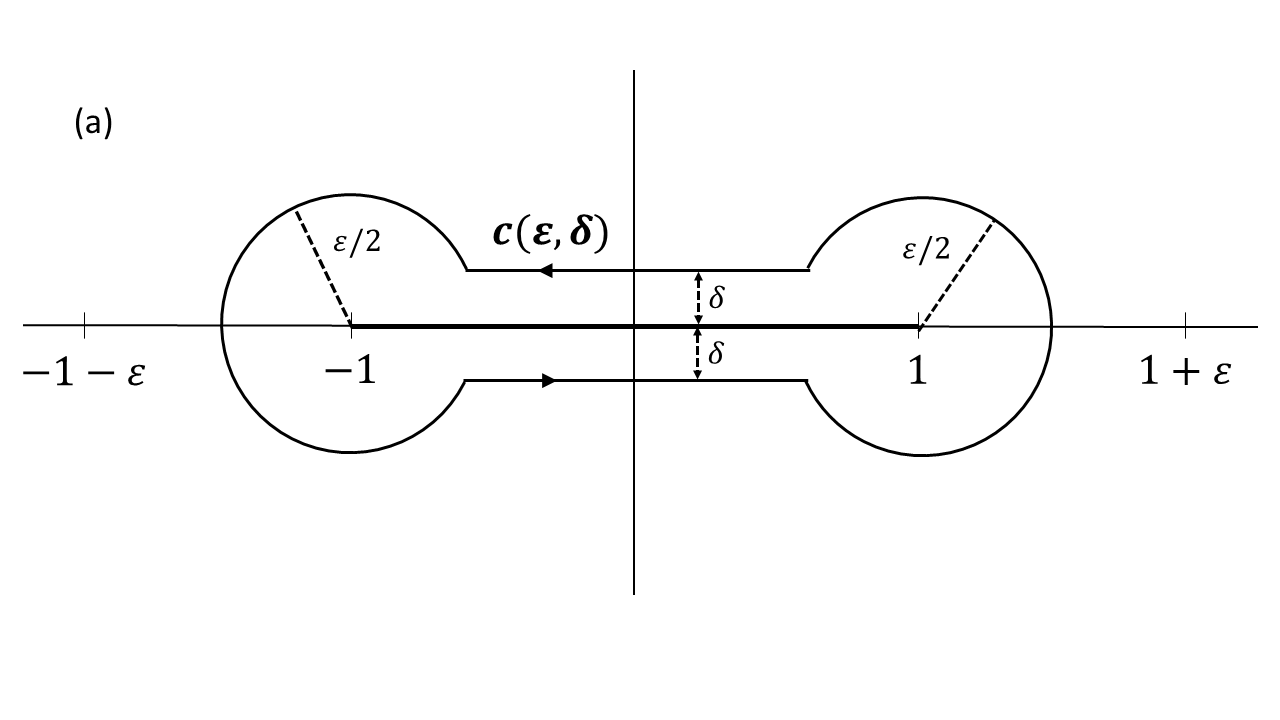}\\
\includegraphics[scale=0.25]{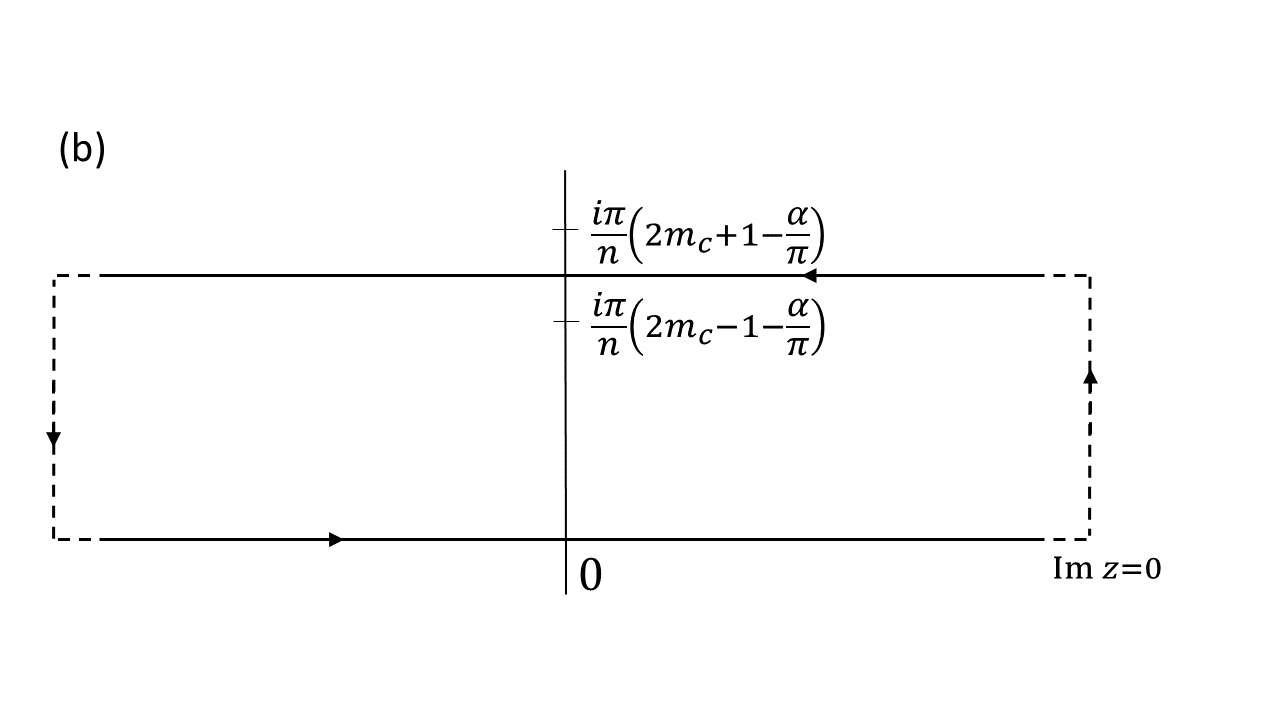}\includegraphics[scale=0.25]{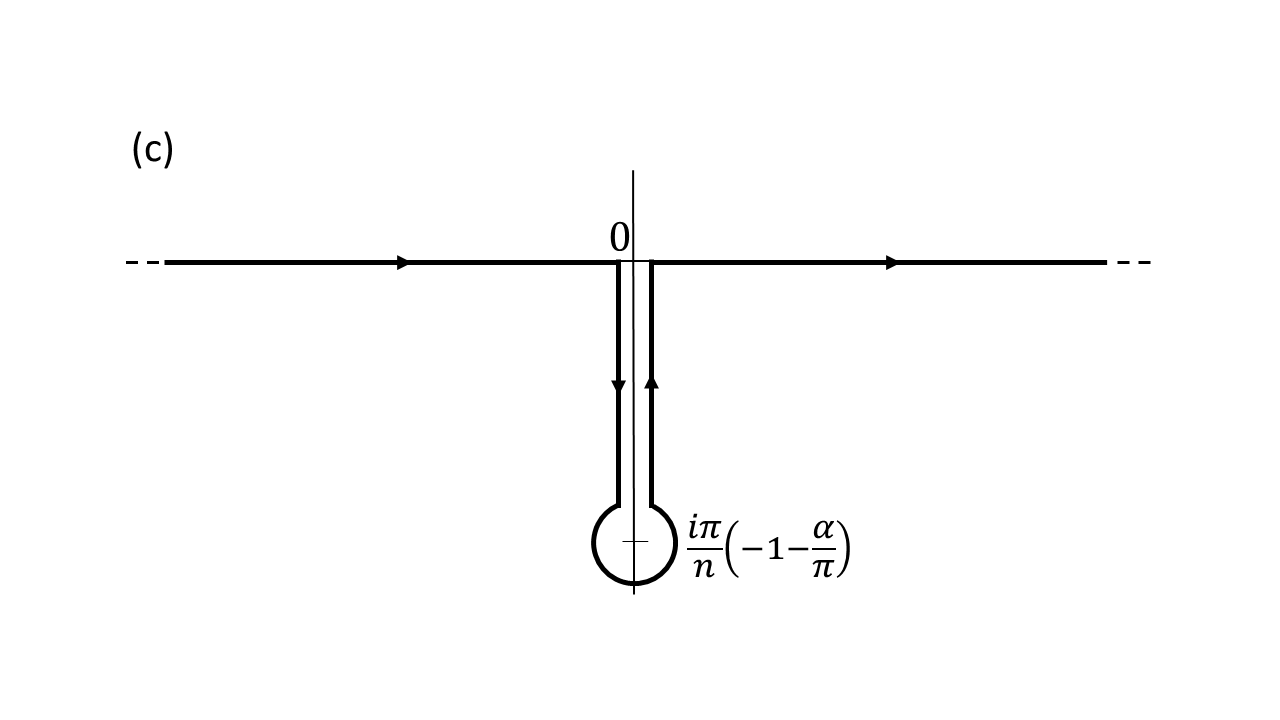}\caption{\textcolor{blue}{\label{fig:The-integration-contour}}(a) The integration
contour $c\left(\varepsilon,\delta\right)$ used in (\ref{eq: Contour Integral}).
(b) The integration contour for the calculation of $I_{k}^{+}$ in
(\ref{eq: definition of I_k^+}). The broken vertical lines represent
segments which are infinitely far from the imaginary line. (c) The
deformed integration contour used in the calculation of $\Upsilon_{0,a}\left(n,\alpha+2\pi\right)$
in (\ref{eq: 2pi shift of upsilon_0}).}
\end{figure}

Following \cite{2004JSP...116...79J}, we calculate $\ln S_{n}\left(\alpha\right)$
for $-\pi<\alpha<\pi$ using integration in the complex plane. We
write
\begin{equation}
\ln S_{n}\left(\alpha\right)=i\frac{\alpha}{2}L+\underset{m=1}{\overset{L}{\sum}}e_{n}^{\left(\alpha\right)}\left(1,\nu_{m}\right)=i\frac{\alpha}{2}L+\underset{\varepsilon,\delta\rightarrow0^{+}}{\lim}\frac{1}{2\pi i}\underset{c\left(\varepsilon,\delta\right)}{\int}e_{n}^{\left(\alpha\right)}\left(1+\varepsilon,\lambda\right)\frac{d}{d\lambda}\ln D_{L}\left(\lambda\right)d\lambda,\label{eq: Contour Integral}
\end{equation}
where $D_{L}\left(\lambda\right)\equiv\det\left(\lambda I_{L}-G_{L}\right)$
as before, $c\left(\varepsilon,\delta\right)$ is the contour presented
in Fig. \ref{fig:The-integration-contour}(a), and
\begin{equation}
e_{n}^{\left(\alpha\right)}\left(x,\nu\right)\equiv\ln\left[\left(\frac{x+\nu}{2}\right)^{n}e^{i\frac{\alpha}{2}}+\left(\frac{x-\nu}{2}\right)^{n}e^{-i\frac{\alpha}{2}}\right].
\end{equation}

We begin by omitting subleading contributions to the asymptotic expression
for $D_{L}\left(\lambda\right)$, substituting for it the leading
order approximation (\ref{eq: FH leading}). We will accordingly obtain
a leading order approximation for $\ln S_{n}\left(\alpha\right)$
at large $L$; this approximation will be denoted by $\ln S_{n}^{\left(0\right)}\left(\alpha\right)$.
One can show that
\begin{equation}
\frac{d}{d\lambda}\ln D_{L}^{\left(0\right)}\left(\lambda\right)=\left(\frac{k_{F}/\pi}{\lambda-1}+\frac{1-k_{F}/\pi}{\lambda+1}\right)L-\frac{4i}{\pi}\cdot\frac{\beta\left(\lambda\right)}{\left(\lambda+1\right)\left(\lambda-1\right)}\left[\ln\mathcal{L}+\left(1+\gamma_{E}\right)+\Upsilon\left(\lambda\right)\right],\label{eq: Fisher Hartwig dlog}
\end{equation}
where 
\begin{equation}
\Upsilon\left(\lambda\right)\equiv\underset{k=1}{\overset{\infty}{\sum}}\frac{k^{-1}\beta^{2}\left(\lambda\right)}{k^{2}-\beta^{2}\left(\lambda\right)}.
\end{equation}
Substituting (\ref{eq: Fisher Hartwig dlog}) into (\ref{eq: Contour Integral})
we get
\begin{align}
\ln S_{n}^{\left(0\right)}\left(\alpha\right) & =i\frac{\alpha}{2}L+\underset{\varepsilon,\delta\rightarrow0^{+}}{\lim}\frac{1}{2\pi i}\underset{c\left(\varepsilon,\delta\right)}{\int}e_{n}^{\left(\alpha\right)}\left(1+\varepsilon,\lambda\right)\left(\frac{k_{F}/\pi}{\lambda-1}+\frac{1-k_{F}/\pi}{\lambda+1}\right)Ld\lambda+\nonumber \\
 & +\underset{\varepsilon,\delta\rightarrow0^{+}}{\lim}\frac{1}{2\pi i}\underset{c\left(\varepsilon,\delta\right)}{\int}e_{n}^{\left(\alpha\right)}\left(1+\varepsilon,\lambda\right)\left(-\frac{4i}{\pi}\cdot\frac{\beta\left(\lambda\right)}{\left(\lambda+1\right)\left(\lambda-1\right)}\left[\ln\mathcal{L}+\left(1+\gamma_{E}\right)+\Upsilon\left(\lambda\right)\right]\right)d\lambda.\label{eq: Contour int. with FH dlog}
\end{align}
Calculating the integrals, we obtain
\begin{equation}
\ln S_{n}^{\left(0\right)}\left(\alpha\right)=i\frac{k_{F}}{\pi}\alpha L+\left[\frac{1}{6}\left(\frac{1}{n}-n\right)-\frac{\alpha^{2}}{2\pi^{2}n}\right]\ln\mathcal{L}+\Upsilon_{0}\left(n,\alpha\right)\,\,\,\left(-\pi<\alpha<\pi\right),\label{eq: 0th order expression}
\end{equation}
where we have defined
\begin{equation}
\Upsilon_{0}\left(n,\alpha\right)\equiv-\frac{1}{\pi^{2}}\underset{0}{\overset{\infty}{\int}}\ln\left[\frac{2\cos\alpha+2\cosh\left(nu\right)}{\left(2\cosh\left(\frac{u}{2}\right)\right)^{2n}}\right]du\underset{0}{\overset{\infty}{\int}}\left[\frac{e^{-t}}{t}-\frac{\cos\left(\frac{ut}{2\pi}\right)}{2\sinh\left(\frac{t}{2}\right)}\right]dt.
\end{equation}
 An equivalent expression for $\Upsilon_{0}\left(n,\alpha\right)$,
which will be of use later on, is
\begin{equation}
\Upsilon_{0}\left(n,\alpha\right)=\frac{in}{2\pi}\underset{-\infty}{\overset{\infty}{\int}}\left[\tanh\left(\frac{nu}{2}+i\frac{\alpha}{2}\right)-\tanh\left(\frac{u}{2}\right)\right]\ln\frac{\Gamma\left(\frac{1}{2}+\frac{u}{2\pi i}\right)}{\Gamma\left(\frac{1}{2}-\frac{u}{2\pi i}\right)}du.\label{eq: Upsilon ver2}
\end{equation}
It is important to note that the $\alpha^{2}$ term in (\ref{eq: 0th order expression})
arises from a Fourier series, $\alpha^{2}=\frac{\pi^{2}}{3}+4\underset{k=1}{\overset{\infty}{\sum}}\frac{\left(-1\right)^{k}}{k^{2}}\cos\left(k\alpha\right)$,
and therefore it should actually be continued periodically outside
the interval $\left[-\pi,\pi\right]$. The calculation of (\ref{eq: 0th order expression})
is detailed in subsection \ref{subsec: appendix 1} of the appendix.

It is noteworthy that the term $\Upsilon_{0}\left(n,\alpha\right)$
is independent of $L$ and $k_{F}$, and that it is real and even
with respect to $\alpha$. We can therefore write
\begin{equation}
\Upsilon_{0}\left(n,\alpha\right)=c_{0}\left(n\right)+c_{2}\left(n\right)\alpha^{2}+\mathcal{O}\left(\alpha^{4}\right).
\end{equation}
Knowing the values $c_{0}\left(n\right)$ and $c_{2}\left(n\right)$
lets us write $\ln S_{n}^{\left(0\right)}\left(\alpha\right)$ as
a quadratic polynomial in $\alpha$:
\begin{equation}
\ln S_{n}^{\left(0\right)}\left(\alpha\right)\approx c_{0}\left(n\right)+\frac{1}{6}\left(\frac{1}{n}-n\right)\ln\mathcal{L}+i\frac{k_{F}}{\pi}L\alpha-\frac{1}{2}\left(\frac{\ln\mathcal{L}}{\pi^{2}n}-2c_{2}\left(n\right)\right)\alpha^{2}\equiv\ln S_{n}^{\left(G\right)}\left(\alpha\right),\label{eq: flux-resolved XX gaussian}
\end{equation}
In such a way the flux-resolved REE is approximated (up to a phase
and a normalization constant) as a density function of a Gaussian
distribution $S_{n}^{\left(G\right)}\left(\alpha\right)$, which implies
that under this approximation its Fourier transform --- the charge-resolved
REE --- represents a Gaussian distribution as well:
\begin{equation}
S_{n}\left(Q_{A}\right)\approx e^{c_{0}\left(n\right)}{\cal L}^{\frac{1}{6}\left(\frac{1}{n}-n\right)}\sqrt{\frac{1}{\frac{2\ln\mathcal{L}}{\pi n}-4\pi c_{2}\left(n\right)}}\exp\left[-\frac{\pi\left(Q_{A}-\frac{k_{F}}{\pi}L\right)^{2}}{\frac{2\ln\mathcal{L}}{\pi n}-4\pi c_{2}\left(n\right)}\right].\label{eq: charge-resolved XX gaussian}
\end{equation}
The deviation of $S_{n}^{\left(G\right)}\left(\alpha\right)$
from $S_{n}^{\left(0\right)}\left(\alpha\right)$ is obviously small
as long as $\left|\alpha\right|\ll\pi$. If we demand that $\ln\mathcal{L}\slash n\gg1$,
subleading corrections to $S_{n}^{\left(0\right)}\left(\alpha\right)$
do not spoil this (for $\left|\alpha\right|\ll\pi$ these subleading
corrections, which we obtain below, vanish exponentially as $\ln\mathcal{L}\slash n\rightarrow\infty$),
meaning that $S_{n}\left(\alpha\right)\approx S_{n}^{\left(G\right)}\left(\alpha\right)$
constitutes a decent approximation in the $\left|\alpha\right|\ll\pi$
regime. Furthermore, the condition $\ln\mathcal{L}\slash n\gg1$ guarantees
that the main contribution to the integral in (\ref{eq: flux-resolved to charge-resolved})
will come from the $\left|\alpha\right|\ll\pi$ regime, due to the
fast decay of the $\exp\left[-\frac{1}{2}\left(\frac{\ln\mathcal{L}}{\pi^{2}n}-2c_{2}\left(n\right)\right)\alpha^{2}\right]$
term away from $\alpha=0$. We can therefore deduce that the Gaussian
approximation (\ref{eq: charge-resolved XX gaussian}) is valid as
long as $\ln\mathcal{L}\slash n\gg1$. We will test the quality of
this approximation in the next subsection.

The value of $c_{2}\left(n\right)$ for the case $n=1$ is of special
interest: since $S_{1}\left(Q_{A}\right)$ is the charge distribution
in subsystem $A$, the expression $\left(\frac{\ln\mathcal{L}}{\pi^{2}}-2c_{2}\left(1\right)\right)$
corresponds to the charge variance. Substituting $n=1$, the value
$c_{2}\left(1\right)=-\frac{1+\gamma_{E}}{2\pi^{2}}$ is obtained
(a detailed proof is presented in subsection \ref{subsec: appendix 2}
of the appendix). This agrees with \cite{PhysRevB.85.035409}, where
it was proven that for a half-filled chain ($k_{F}=\frac{\pi}{2}$,
and accordingly $\mathcal{L}=2L$) the charge variance is $\frac{\ln2L+1+\gamma_{E}}{\pi^{2}}$. 

\subsection{Corrections up to the order of ${\cal O}\left(\mathcal{L}^{-1}\right)$}

Corrections to the leading order approximation (\ref{eq: 0th order expression})
can be calculated by taking into account subleading contributions
that appear in (\ref{eq: FH subleading}). Following \cite{Calabrese_2010},
we use the fact that $G\left(1+x\right)/G\left(x\right)=\Gamma\left(x\right)$
and, omitting terms which will contribute corrections of order $\mathcal{O}\left(\mathcal{L}^{-4}\right)$,
we rewrite (\ref{eq: FH subleading}) as
\begin{align}
D_{L}\left(\lambda\right) & =D_{L}^{\left(0\right)}\left(\lambda\right)\left[1+e^{2ik_{F}L}\mathcal{L}^{-2+4\beta\left(\lambda\right)}\frac{\Gamma\left(1-\beta\left(\lambda\right)\right)^{2}}{\Gamma\left(\beta\left(\lambda\right)\right)^{2}}+e^{-2ik_{F}L}\mathcal{L}^{-2-4\beta\left(\lambda\right)}\frac{\Gamma\left(1+\beta\left(\lambda\right)\right)^{2}}{\Gamma\left(-\beta\left(\lambda\right)\right)^{2}}\right]\equiv\nonumber \\
 & \equiv D_{L}^{\left(0\right)}\left(\lambda\right)\left[1+H\left(\lambda\right)\right].
\end{align}
Substituting this into the integral expression for $\ln S_{n}\left(\alpha\right)$
(\ref{eq: Contour Integral}), we obtain 
\begin{align}
\ln S_{n}\left(\alpha\right) & =\ln S_{n}^{\left(0\right)}\left(\alpha\right)+\underset{\varepsilon,\delta\rightarrow0^{+}}{\lim}\frac{1}{2\pi i}\underset{c\left(\varepsilon,\delta\right)}{\int}e_{n}^{\left(\alpha\right)}\left(1+\varepsilon,\lambda\right)\frac{d}{d\lambda}\ln\left[1+H\left(\lambda\right)\right]d\lambda+\mathcal{O}\left(\mathcal{L}^{-4}\right)=\nonumber \\
 & =\ln S_{n}^{\left(0\right)}\left(\alpha\right)-\underset{\varepsilon,\delta\rightarrow0^{+}}{\lim}\frac{1}{2\pi i}\underset{c\left(\varepsilon,\delta\right)}{\int}\frac{de_{n}^{\left(\alpha\right)}\left(1+\varepsilon,\lambda\right)}{d\lambda}\ln\left[1+H\left(\lambda\right)\right]d\lambda+\mathcal{O}\left(\mathcal{L}^{-4}\right).
\end{align}
Using the fact that for every $-1<x<1$,
\begin{equation}
\beta\left(x+i0^{\pm}\right)=-iW\left(x\right)\mp\frac{1}{2},
\end{equation}
where $W\left(x\right)\equiv\frac{1}{2\pi}\ln\frac{1+x}{1-x}$, we
obtain that
\begin{equation}
H\left(x+i0^{\pm}\right)=e^{\pm2i\left(2\ln\mathcal{L}W\left(x\right)-k_{F}L\right)}\frac{\Gamma\left(\frac{1}{2}\mp iW\left(x\right)\right)^{2}}{\Gamma\left(\frac{1}{2}\pm iW\left(x\right)\right)^{2}}+\mathcal{O}\left(\mathcal{L}^{-4}\right).
\end{equation}
Now we write $\ln\left[1+H\left(\lambda\right)\right]=\underset{k=1}{\overset{\infty}{\sum}}\frac{\left(-1\right)^{k+1}}{k}H\left(\lambda\right)^{k}$
and take the limit $\varepsilon,\delta\rightarrow0^{+}$, omitting
terms of order $\mathcal{O}\left(\mathcal{L}^{-4}\right)$, so that
we get
\begin{align}
\ln S_{n}\left(\alpha\right)-\ln S_{n}^{\left(0\right)}\left(\alpha\right) & =\frac{1}{2\pi i}\underset{-\infty}{\overset{\infty}{\int}}du\underset{k=1}{\overset{\infty}{\sum}}\frac{\left(-1\right)^{k+1}n}{2k}\left[\tanh\left(\frac{nu}{2}+i\frac{\alpha}{2}\right)-\tanh\left(\frac{u}{2}\right)\right]\times\nonumber \\
 & \times\left\{ e^{2ik\left(\frac{\ln\mathcal{L}}{\pi}u-k_{F}L\right)}\frac{\Gamma\left(\frac{1}{2}+\frac{u}{2\pi i}\right)^{2k}}{\Gamma\left(\frac{1}{2}-\frac{u}{2\pi i}\right)^{2k}}-e^{-2ik\left(\frac{\ln\mathcal{L}}{\pi}u-k_{F}L\right)}\frac{\Gamma\left(\frac{1}{2}-\frac{u}{2\pi i}\right)^{2k}}{\Gamma\left(\frac{1}{2}+\frac{u}{2\pi i}\right)^{2k}}\right\} .
\end{align}

Let us define a natural number $m_{c}=m_{c}\left(n\right)\equiv\lceil\frac{n}{4}\rceil+1$,
so that $m_{c}\ge\frac{n}{4}+1$, and thus $\frac{2}{n}\left(2m_{c}-1\pm\frac{\alpha}{\pi}\right)\ge1$
for every $-\pi<\alpha<\pi$. For each $k\ge1$ and every $-\pi<\alpha<\pi$,
we can estimate the integral
\begin{equation}
I_{k}^{+}\equiv\frac{1}{2\pi i}\underset{-\infty}{\overset{\infty}{\int}}du\frac{\left(-1\right)^{k+1}n}{2k}\left[\tanh\left(\frac{nu}{2}+i\frac{\alpha}{2}\right)-\tanh\left(\frac{u}{2}\right)\right]e^{2ik\left(\frac{\ln\mathcal{L}}{\pi}u-k_{F}L\right)}\frac{\Gamma\left(\frac{1}{2}+\frac{u}{2\pi i}\right)^{2k}}{\Gamma\left(\frac{1}{2}-\frac{u}{2\pi i}\right)^{2k}}\label{eq: definition of I_k^+}
\end{equation}
by enclosing the $m_{c}$ poles of $\tanh\left(\frac{nz}{2}+i\frac{\alpha}{2}\right)$
that are in the upper half-plane (at $z=\frac{i\pi}{n}\left(2m-1-\frac{\alpha}{\pi}\right)$
for $m\in\mathbb{N}$) and are closest to the real line using a rectangular
contour, the vertical sides of which are infinitely far from the imaginary
line, and whose upper horizontal side crosses the imaginary line through
the segment between the $m_{c}$-th and the $\left(m_{c}+1\right)$-th
pole of $\tanh\left(\frac{nz}{2}+i\frac{\alpha}{2}\right)$ (see Fig.
\ref{fig:The-integration-contour}(b)). Thus we make sure that the
integral over the upper horizontal side of the contour is of order
$\mathcal{O}\left(\mathcal{L}^{-1-\frac{4}{n}}\right)$
at most. Ignoring the poles of $\tanh\left(\frac{z}{2}\right)$, considering
that the contribution of their residues is only of order $\mathcal{O}\left(\mathcal{L}^{-2}\right)$,
we can write
\begin{equation}
I_{k}^{+}=\underset{m=1}{\overset{m_{c}}{\sum}}\frac{\left(-1\right)^{k+1}}{k}\mathcal{L}^{-\frac{2k}{n}\left(2m-1-\frac{\alpha}{\pi}\right)}e^{-2ik_{F}Lk}\frac{\Gamma\left(\frac{1}{2}+\frac{1}{2n}\left(2m-1-\frac{\alpha}{\pi}\right)\right)^{2k}}{\Gamma\left(\frac{1}{2}-\frac{1}{2n}\left(2m-1-\frac{\alpha}{\pi}\right)\right)^{2k}}+\mathcal{O}\left(\mathcal{L}^{-1-\frac{4}{n}}+\mathcal{L}^{-2}\right).
\end{equation}
In a similar way, we define
\begin{equation}
I_{k}^{-}\equiv-\frac{1}{2\pi i}\underset{-\infty}{\overset{\infty}{\int}}du\frac{\left(-1\right)^{k+1}n}{2k}\left[\tanh\left(\frac{nu}{2}+i\frac{\alpha}{2}\right)-\tanh\left(\frac{u}{2}\right)\right]e^{-2ik\left(\frac{\ln\mathcal{L}}{\pi}u-k_{F}L\right)}\frac{\Gamma\left(\frac{1}{2}-\frac{u}{2\pi i}\right)^{2k}}{\Gamma\left(\frac{1}{2}+\frac{u}{2\pi i}\right)^{2k}},
\end{equation}
and sum over the residues of $\tanh\left(\frac{nz}{2}+i\frac{\alpha}{2}\right)$
at its poles in the lower half-plane up to $m=m_{c}$, so that we
get
\begin{equation}
I_{k}^{-}=\underset{m=1}{\overset{m_{c}}{\sum}}\frac{\left(-1\right)^{k+1}}{k}\mathcal{L}^{-\frac{2k}{n}\left(2m-1+\frac{\alpha}{\pi}\right)}e^{2ik_{F}Lk}\frac{\Gamma\left(\frac{1}{2}+\frac{1}{2n}\left(2m-1+\frac{\alpha}{\pi}\right)\right)^{2k}}{\Gamma\left(\frac{1}{2}-\frac{1}{2n}\left(2m-1+\frac{\alpha}{\pi}\right)\right)^{2k}}+\mathcal{O}\left(\mathcal{L}^{-1-\frac{4}{n}}+\mathcal{L}^{-2}\right).
\end{equation}
The summation over $m$ is truncated due the fact that infinite summation
will not converge.{} Further corrections of order $o\left(\mathcal{L}^{-1}\right)$
that are not captured by the generalized Fisher-Hartwig conjecture,
and stem from a calculation related to random matrix theory, were
found in the calculation of the total REE in \cite{Calabrese_2010}.

Summing over $k$, we finally get
\begin{equation}
\ln S_{n}\left(\alpha\right)=\ln S_{n}^{\left(0\right)}\left(\alpha\right)+\Upsilon_{1}\left(n,\alpha,L,k_{F}\right)+o\left(\mathcal{L}^{-1}\right)\,\,\,\left(-\pi<\alpha<\pi\right),\label{eq: 1st order expression}
\end{equation}
where
\begin{align}
\Upsilon_{1}\left(n,\alpha,L,k_{F}\right) & \equiv\underset{m=1}{\overset{m_{c}}{\sum}}\ln\left[1+\mathcal{L}^{-\frac{2}{n}\left(2m-1-\frac{\alpha}{\pi}\right)}e^{-2ik_{F}L}\frac{\Gamma\left(\frac{1}{2}+\frac{1}{2n}\left(2m-1-\frac{\alpha}{\pi}\right)\right)^{2}}{\Gamma\left(\frac{1}{2}-\frac{1}{2n}\left(2m-1-\frac{\alpha}{\pi}\right)\right)^{2}}\right]+\nonumber \\
 & +\underset{m=1}{\overset{m_{c}}{\sum}}\ln\left[1+\mathcal{L}^{-\frac{2}{n}\left(2m-1+\frac{\alpha}{\pi}\right)}e^{2ik_{F}L}\frac{\Gamma\left(\frac{1}{2}+\frac{1}{2n}\left(2m-1+\frac{\alpha}{\pi}\right)\right)^{2}}{\Gamma\left(\frac{1}{2}-\frac{1}{2n}\left(2m-1+\frac{\alpha}{\pi}\right)\right)^{2}}\right].
\end{align}

\begin{figure}[h]
\begin{centering}
\includegraphics[scale=0.5]{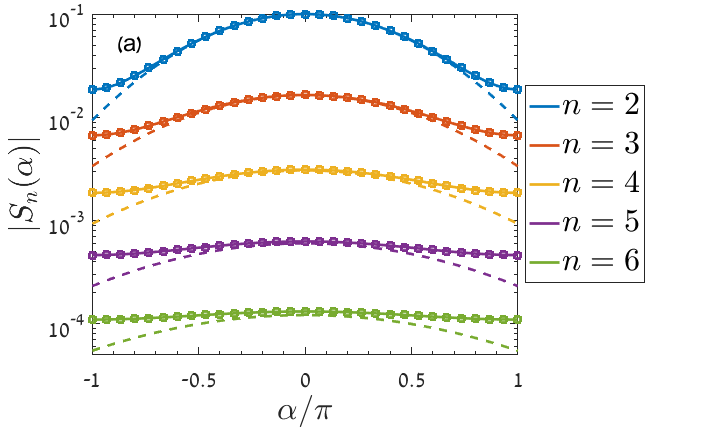}\includegraphics[scale=0.5]{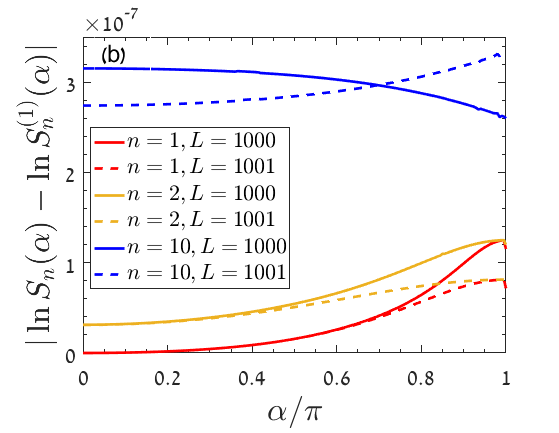}\\
\par\end{centering}
\caption{\label{fig:Flux-resolved-REE-in}(a) Flux-resolved REE in a subsystem
of length $L=1000$ of a half-filled gapless XX chain, computed numerically
according to (\ref{eq: flux-resolved REE explicit}) (dots), using
the analytical leading order approximation (\ref{eq: 0th order expression})
(broken lines), and using the analytical approximation up to ${\cal O}\left({\cal L}^{-1}\right)$
(\ref{eq: 1st order expression}) (continuous lines). (b) The absolute
deviation of the analytical result up to order $\mathcal{O}\left(\mathcal{L}^{-1}\right)$
from the numerical result for the flux-resolved REE, for a half-filled
gapless XX chain.}
\end{figure}

Fig. \ref{fig:Flux-resolved-REE-in}(a) shows the dependence of the
flux-resolved REE on $\alpha$ in a half-filled system ($k_{F}=\frac{\pi}{2}$),
for different values of $n$. The numerical evaluation of (\ref{eq: flux-resolved REE explicit})
is compared to the analytical results, and it can be seen that while
the leading order approximation $S_{n}^{\left(0\right)}\left(\alpha\right)$
exhibits an $\mathcal{O}\left(1\right)$ deviation from the numerical
values as $\alpha\rightarrow\pm\pi$, this deviation practically vanishes
after we include corrections up to order ${\cal O}\left({\cal L}^{-1}\right)$.
Fig. \ref{fig:Flux-resolved-REE-in}(b) shows a more detailed comparison
between the analytical result up to order $\mathcal{O}\left(\mathcal{L}^{-1}\right)$
and the numerical result, for the case of half-filling. In this figure
we denote the analytical result by $\ln S_{n}^{\left(1\right)}\left(\alpha\right)\equiv\ln S_{n}^{\left(0\right)}\left(\alpha\right)+\Upsilon_{1}\left(n,\alpha,L,k_{F}\right)$,
while the numerical result is denoted by $\ln S_{n}\left(\alpha\right)$.
The negligible difference between the two calculations indicates that
$\ln S_{n}^{\left(1\right)}\left(\alpha\right)$ provides a very good
approximation even for a subsystem of relatively moderate length.

\begin{figure}[h]
\centering{}\includegraphics[scale=0.5]{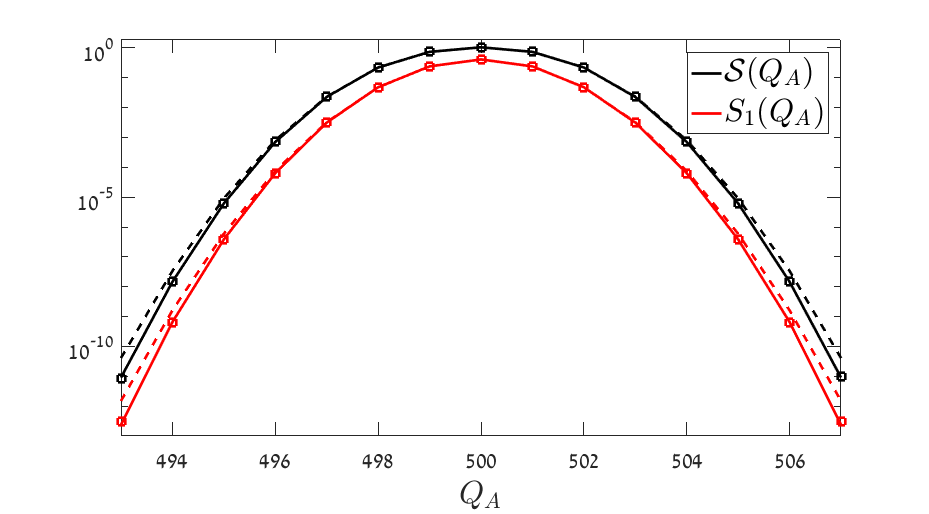}\caption{\label{fig: charge resolved XX}Charge-resolved REE and vNEE in a
subsystem of length $L=1000$ of a half-filled gapless XX chain, computed
numerically according to (\ref{eq: flux-resolved REE explicit}) and
(\ref{eq: flux-resolved to charge-resolved}) (dots), using the analytical
Gaussian approximation (\ref{eq: charge-resolved XX gaussian}) (broken
lines), and using the analytical approximation up to ${\cal O}\left({\cal L}^{-1}\right)$
according to (\ref{eq: 1st order expression}) and (\ref{eq: flux-resolved to charge-resolved})
(continuous lines).}
\end{figure}

We can now use the analytical results for $S_{n}\left(\alpha\right)$
in order to calculate the charge-resolved REE through (\ref{eq: flux-resolved to charge-resolved}),
and then the charge-resolved vNEE. Fig. \ref{fig: charge resolved XX}
shows that when we use the analytical approximation $S_{n}^{\left(1\right)}\left(\alpha\right)$,
these calculations are in good agreement with numerical results. On
the other hand, the Gaussian approximation derived from (\ref{eq: charge-resolved XX gaussian})
exhibits a discernible deviation from numerical results for both $S_{1}\left(Q_{A}\right)$
and ${\cal S}\left(Q_{A}\right)$, since $\ln\mathcal{L}$ is not
large enough.

\subsection{Periodic structure up to the order of ${\cal O}\left(\mathcal{L}^{-1}\right)$\label{subsec: Periodic structure}}

$\ln S_{n}^{\left(0\right)}\left(\alpha\right)$ in (\ref{eq: 0th order expression})
was defined for $-\pi<\alpha<\pi$, and its real part is $2\pi$-periodic
in $\alpha$ (remember that the $\alpha^{2}$ term originated from
a Fourier series). Nevertheless, its periodic continuation is not
analytic (nor is it even differentiable), so we would like to define
the analytic continuation of $S_{n}^{\left(0\right)}\left(\alpha\right)$
for $\alpha\in\mathbb{R}$. For this purpose, we will construct a
natural continuation of $\ln S_{n}^{\left(0\right)}\left(\alpha\right)$
so that the corresponding continuation of $S_{n}^{\left(0\right)}\left(\alpha\right)$,
which will be denoted by $S_{a,n}^{\left(0\right)}\left(\alpha\right)$,
would turn out analytic. The linear and quadratic terms in (\ref{eq: 0th order expression})
naturally remain as before, so we need only to construct an appropriate
continuation $\Upsilon_{0,a}\left(n,\alpha\right)$ of the term $\Upsilon_{0}\left(n,\alpha\right)$,
and then obtain for $\alpha\in\mathbb{R}$
\begin{equation}
S_{a,n}^{\left(0\right)}\left(\alpha\right)=\exp\left\{ i\frac{k_{F}}{\pi}\alpha L+\left[\frac{1}{6}\left(\frac{1}{n}-n\right)-\frac{\alpha^{2}}{2\pi^{2}n}\right]\ln\mathcal{L}+\Upsilon_{0,a}\left(n,\alpha\right)\right\} .\label{eq: Renyi Analytic Continuation}
\end{equation}

Regarding the term $\Upsilon_{0}\left(n,\alpha\right)$ as it is written
in (\ref{eq: Upsilon ver2}), note that as $\alpha$ approaches $\pi^{-}$
or $-\pi^{+}$, a pole of the function $\tanh\left(\frac{nz}{2}+i\frac{\alpha}{2}\right)$
approaches the real line. A shift of $\alpha\rightarrow\alpha+2\pi$
maintains the positions of all poles of $\tanh\left(\frac{nz}{2}+i\frac{\alpha}{2}\right)$
in the upper half-plane (at $z=\frac{i\pi}{n}\left(2m-1-\frac{\alpha}{\pi}\right),\,m\in\mathbb{N}$),
but during a continuous shift of such kind the pole that was originally
at $z=\frac{i\pi}{n}\left(1-\frac{\alpha}{\pi}\right)$ crosses the
real line, and ends up at $z=\frac{i\pi}{n}\left(-1-\frac{\alpha}{\pi}\right)$.
We can now think of $\Upsilon_{0,a}\left(n,\alpha+2\pi\right)$ as
the value obtained by calculating the integral in (\ref{eq: Upsilon ver2})
while deforming the contour of integration (originally just the real
line) so that it also encircles the pole that crossed the real line,
thus counting the residue of the integrand at $z=\frac{i\pi}{n}\left(-1-\frac{\alpha}{\pi}\right)$
(see Fig. \ref{fig:The-integration-contour}(c)). In such a way we
get for every $-\pi<\alpha<\pi$,
\begin{align}
\Upsilon_{0,a}\left(n,\alpha+2\pi\right)-\Upsilon_{0}\left(n,\alpha\right) & =\text{Res}\left\{ -n\left[\tanh\left(\frac{nz}{2}+i\frac{\alpha}{2}\right)-\tanh\left(\frac{z}{2}\right)\right]\ln\frac{\Gamma\left(\frac{1}{2}+\frac{z}{2\pi i}\right)}{\Gamma\left(\frac{1}{2}-\frac{z}{2\pi i}\right)},z=\frac{i\pi}{n}\left(-1-\frac{\alpha}{\pi}\right)\right\} =\nonumber \\
 & =2\ln\frac{\Gamma\left(\frac{1}{2}+\frac{1}{2n}\left(1+\frac{\alpha}{\pi}\right)\right)}{\Gamma\left(\frac{1}{2}-\frac{1}{2n}\left(1+\frac{\alpha}{\pi}\right)\right)}.\label{eq: 2pi shift of upsilon_0}
\end{align}
By the same logic, for every natural number $m\ge1$ we can deform
the integration contour so that it encircles the $m$ poles which
cross the real line from the upper half-plane to the lower half-plane
during the shift $\alpha\rightarrow\alpha+2\pi m$, so that for every
$-\pi<\alpha<\pi$,
\begin{align}
\Upsilon_{0,a}\left(n,\alpha+2\pi m\right)-\Upsilon_{0}\left(n,\alpha\right) & =\underset{j=1}{\overset{m}{\sum}}\text{Res}\left\{ -n\left[\tanh\left(\frac{nz}{2}+i\frac{\alpha}{2}\right)-\tanh\left(\frac{z}{2}\right)\right]\ln\frac{\Gamma\left(\frac{1}{2}+\frac{z}{2\pi i}\right)}{\Gamma\left(\frac{1}{2}-\frac{z}{2\pi i}\right)},z=\frac{i\pi}{n}\left(-2j+1-\frac{\alpha}{\pi}\right)\right\} =\nonumber \\
 & =\underset{j=1}{\overset{m}{\sum}}2\ln\frac{\Gamma\left(\frac{1}{2}+\frac{1}{2n}\left(2j-1+\frac{\alpha}{\pi}\right)\right)}{\Gamma\left(\frac{1}{2}-\frac{1}{2n}\left(2j-1+\frac{\alpha}{\pi}\right)\right)}.\label{eq: Analytic upsilon positive shift}
\end{align}
For a shift of $\alpha\rightarrow\alpha-2\pi m$ (this time encircling
poles that cross the real line from the lower half-plane to the upper
half-plane), we get for every $-\pi<\alpha<\pi$,
\begin{equation}
\Upsilon_{0,a}\left(n,\alpha-2\pi m\right)-\Upsilon_{0}\left(n,\alpha\right)=\underset{j=1}{\overset{m}{\sum}}2\ln\frac{\Gamma\left(\frac{1}{2}+\frac{1}{2n}\left(2j-1-\frac{\alpha}{\pi}\right)\right)}{\Gamma\left(\frac{1}{2}-\frac{1}{2n}\left(2j-1-\frac{\alpha}{\pi}\right)\right)}.\label{eq: Analytic upsilon negative shift}
\end{equation}

\begin{figure}[h]
\begin{centering}
\includegraphics[scale=0.5]{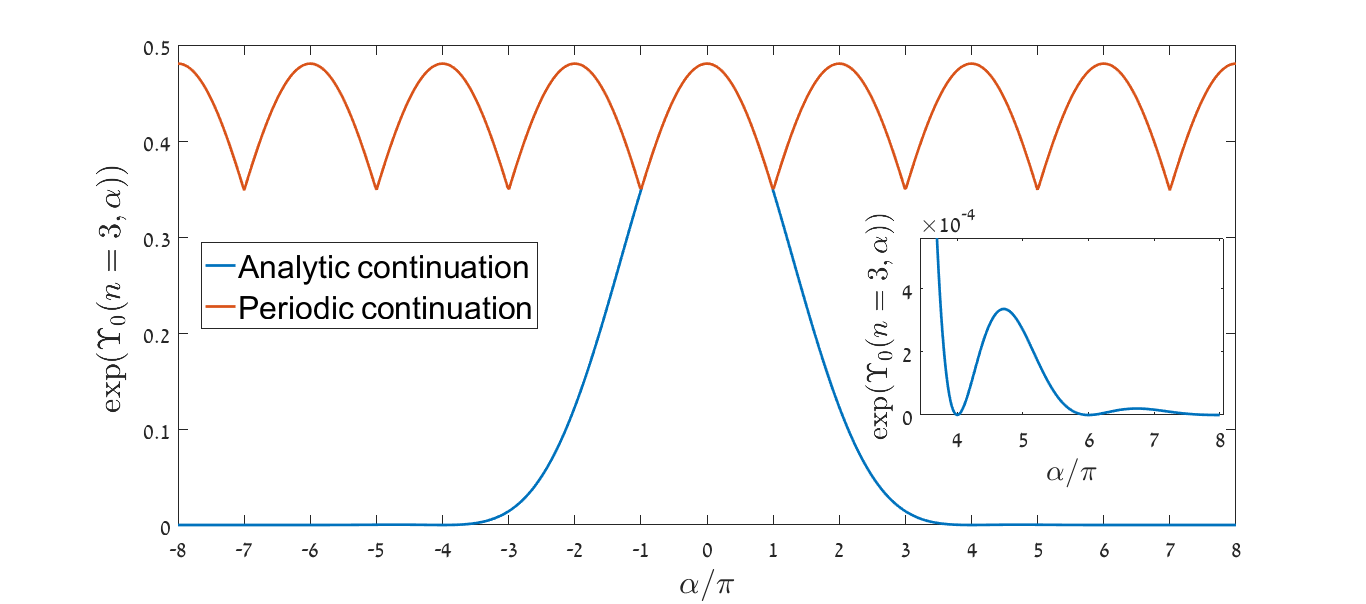}
\par\end{centering}
\caption{\label{fig: Continuation of upsilon}Continuation of $\exp\left(\Upsilon_{0}\left(n,\alpha\right)\right)$
for $n=3$. Inset shows a zoomed-in view of the tail of the analytic
continuation $\exp\left(\Upsilon_{0,a}\left(n=3,\alpha\right)\right)$.}
\end{figure}

For fixed $n$, the terms $\Gamma\left(\frac{1}{2}-\frac{1}{2n}\left(2j-1\pm\frac{\alpha}{\pi}\right)\right)$
might diverge for certain values of $j$ and $\alpha$, in which case
(\ref{eq: Analytic upsilon positive shift}) or (\ref{eq: Analytic upsilon negative shift})
diverge, respectively. This however does not pose a problem, since
we are eventually interested in the exponents of (\ref{eq: Analytic upsilon positive shift})
and (\ref{eq: Analytic upsilon negative shift}), and when $\Gamma\left(\frac{1}{2}-\frac{1}{2n}\left(2j-1\pm\frac{\alpha}{\pi}\right)\right)$
diverge for some $1\le j\le m$ it just means that $\exp\left(\Upsilon_{0,a}\left(n,\alpha\pm2\pi m\right)\right)=0$,
respectively. Both the periodic and the analytic contintuations of
$\exp\Upsilon_{0}$ are presented in Fig. \ref{fig: Continuation of upsilon}.

Defining $\Upsilon_{0,a}\left(n,\alpha\right)$ this way and substituting
it into the analytic continuation of $S_{n}^{\left(0\right)}\left(\alpha\right)$
in (\ref{eq: Renyi Analytic Continuation}), we obtain for each $m\in\mathbb{N}$
and every $-\pi<\alpha<\pi$,
\begin{equation}
\frac{S_{a,n}^{\left(0\right)}\left(\alpha\pm2\pi m\right)}{S_{n}^{\left(0\right)}\left(\alpha\right)}=\mathcal{L}^{-\frac{2}{n}\left(m^{2}\pm\frac{\alpha}{\pi}m\right)}e^{\pm2imk_{F}L}\underset{j=1}{\overset{m}{\Pi}}\frac{\Gamma\left(\frac{1}{2}+\frac{1}{2n}\left(2j-1\pm\frac{\alpha}{\pi}\right)\right)^{2}}{\Gamma\left(\frac{1}{2}-\frac{1}{2n}\left(2j-1\pm\alpha\right)\right)^{2}},
\end{equation}
and in particular
\begin{align}
\frac{S_{a,n}^{\left(0\right)}\left(\alpha+2\pi m\right)}{S_{a,n}^{\left(0\right)}\left(\alpha+2\pi\left(m-1\right)\right)} & =\mathcal{L}^{-\frac{2}{n}\left(2m-1+\frac{\alpha}{\pi}\right)}e^{2ik_{F}L}\frac{\Gamma\left(\frac{1}{2}+\frac{1}{2n}\left(2m-1+\frac{\alpha}{\pi}\right)\right)^{2}}{\Gamma\left(\frac{1}{2}-\frac{1}{2n}\left(2m-1+\frac{\alpha}{\pi}\right)\right)^{2}},\nonumber \\
\frac{S_{a,n}^{\left(0\right)}\left(\alpha-2\pi m\right)}{S_{a,n}^{\left(0\right)}\left(\alpha-2\pi\left(m-1\right)\right)} & =\mathcal{L}^{-\frac{2}{n}\left(2m-1-\frac{\alpha}{\pi}\right)}e^{-2ik_{F}L}\frac{\Gamma\left(\frac{1}{2}+\frac{1}{2n}\left(2m-1-\frac{\alpha}{\pi}\right)\right)^{2}}{\Gamma\left(\frac{1}{2}-\frac{1}{2n}\left(2m-1-\frac{\alpha}{\pi}\right)\right)^{2}}.\label{eq: 2 pi shift relations}
\end{align}

Let us now define $\sigma_{m}\left(\alpha\right)\equiv S_{a,n}^{\left(0\right)}\left(\alpha+2\pi m\right)$
for every $m\in\mathbb{Z}$ and $-\pi<\alpha<\pi$. We can rewrite
(\ref{eq: 1st order expression}) as
\begin{equation}
\ln S_{n}\left(\alpha\right)=\ln\sigma_{0}\left(\alpha\right)+\underset{m=1}{\overset{m_{c}}{\sum}}\left\{ \ln\left[1+\frac{\sigma_{m}\left(\alpha\right)}{\sigma_{m-1}\left(\alpha\right)}\right]+\ln\left[1+\frac{\sigma_{-m}\left(\alpha\right)}{\sigma_{-m+1}\left(\alpha\right)}\right]\right\} +o\left(\mathcal{L}^{-1}\right),
\end{equation}
and therefore, up to $o\left(\mathcal{L}^{-1}\right)$ corrections,
\begin{equation}
S_{n}\left(\alpha\right)=\left[\underset{m=1}{\overset{m_{c}}{\Pi}}\left(\frac{\sigma_{-m}\left(\alpha\right)}{\sigma_{-m+1}\left(\alpha\right)}+1\right)\right]\sigma_{0}\left(\alpha\right)\left[\underset{m=1}{\overset{m_{c}}{\Pi}}\left(1+\frac{\sigma_{m}\left(\alpha\right)}{\sigma_{m-1}\left(\alpha\right)}\right)\right].\label{eq: Renyi truncated multiple}
\end{equation}
We could have formally represented the result in (\ref{eq: 1st order expression})
as an asymptotic (divergent) series had we not defined the cutoff
index $m_{c}$. Such a representation would have brought us to the
asymptotic (divergent) product
\begin{equation}
S_{n}\left(\alpha\right)=\left[\underset{m=1}{\overset{\infty}{\Pi}}\left(\frac{\sigma_{-m}\left(\alpha\right)}{\sigma_{-m+1}\left(\alpha\right)}+1\right)\right]\sigma_{0}\left(\alpha\right)\left[\underset{m=1}{\overset{\infty}{\Pi}}\left(1+\frac{\sigma_{m}\left(\alpha\right)}{\sigma_{m-1}\left(\alpha\right)}\right)\right],
\end{equation}
which for any arbitrary $j\in\mathbb{Z}$ can be written as
\begin{equation}
S_{n}\left(\alpha\right)=\left[\underset{m=1}{\overset{\infty}{\Pi}}\left(\frac{\sigma_{j-m}\left(\alpha\right)}{\sigma_{j-m+1}\left(\alpha\right)}+1\right)\right]\sigma_{j}\left(\alpha\right)\left[\underset{m=1}{\overset{\infty}{\Pi}}\left(1+\frac{\sigma_{j+m}\left(\alpha\right)}{\sigma_{j+m-1}\left(\alpha\right)}\right)\right].
\end{equation}
This result is just $S_{n}\left(\alpha+2\pi j\right)=S_{n}\left(\alpha\right)$,
as long as we ignore $o\left(\mathcal{L}^{-1}\right)$ corrections
and treat it as an asymptotic product.

Note that the result in (\ref{eq: Renyi truncated multiple}) can
also be written as 
\begin{equation}
S_{n}\left(\alpha\right)=\underset{j=-m_{c}}{\overset{m_{c}}{\sum}}S_{a,n}^{\left(0\right)}\left(\alpha+2\pi j\right)+o\left(1\right),\label{eq: periodic entropy structure}
\end{equation}
a structure which is natural from the CFT perspective. Indeed, there
one writes the flux-resolved entropy $S_{n}\left(\alpha\right)$ as
a correlation function over $n$ copies of space-time of $\mathcal{T}_{\mathcal{V}}=\mathcal{T}\times\mathcal{V}$,
twist fields (appearing in the calculation of the total entropies)
$\mathcal{T}$ modified by fusion of vertex operators $\mathcal{V}$,
which assign a phase $\alpha$ to every particle encircling them \cite{PhysRevLett.120.200602}.
In a bosonized language it can be written in terms of the appropriate
boson field $\phi$ as $\mathcal{V}_{0}\left(\alpha\right)=e^{i\frac{\alpha}{2\pi}\phi}$.
However, the periodicity in $\alpha$ implies that $\mathcal{V}$
could actually be taken as a sum over all possible shifts of $\alpha$
by integer multiples of $2\pi$, that is
\begin{equation}
\mathcal{T}_{\mathcal{V}}=\sum_{j}a_{j}(n,\alpha)\mathcal{T}\times\mathcal{V}_{0}(\alpha+2\pi j),
\end{equation}
with some coefficients $a_{j}(n,\alpha)$. Computing the entropies
as in \cite{PhysRevLett.120.200602} would then lead to the form of
(\ref{eq: periodic entropy structure}). Our exact results allow one
to go beyond CFT and find the coefficients for the XX system, which
take the values $a_{j}\left(n,\alpha\right)=\exp\Upsilon_{0,a}\left(n,\alpha+2\pi j\right)$.

Interestingly, this structure is maintained even when we include all
terms up to an order of $\mathcal{O}\left(\mathcal{L}^{-1}\right)$.
Let us define for every $-\pi<\alpha<\pi$
\begin{equation}
\sigma_{\mathrm{right}}\left(\alpha\right)\equiv\sigma_{0}\left(\alpha\right)\left[\underset{m=2}{\overset{m_{c}}{\Pi}}\left(1+\frac{\sigma_{m}\left(\alpha\right)}{\sigma_{m-1}\left(\alpha\right)}\right)\right].
\end{equation}
First, note that (\ref{eq: Renyi truncated multiple}) can be also
written as
\begin{equation}
S_{n}\left(\alpha\right)=\sigma_{-m_{c}}\left(\alpha\right)\left[\underset{m=-m_{c}+1}{\overset{m_{c}}{\Pi}}\left(1+\frac{\sigma_{m}\left(\alpha\right)}{\sigma_{m-1}\left(\alpha\right)}\right)\right]=\underset{j=-m_{c}}{\overset{m_{c}}{\sum}}\sigma_{j}\left(\alpha\right)\underset{m=j+2}{\overset{m_{c}}{\Pi}}\left(1+\frac{\sigma_{m}\left(\alpha\right)}{\sigma_{m-1}\left(\alpha\right)}\right).\label{eq: Renyi rewritten}
\end{equation}
By definition of $m_{c}$, for any $m>m_{c}$ and every $-\pi<\alpha<\pi$
it is true that $\frac{\sigma_{m}\left(\alpha\right)}{\sigma_{m-1}\left(\alpha\right)}=o\left(\mathcal{L}^{-1}\right)$,
and therefore for every $0\le j\le m_{c}$,
\begin{align}
\sigma_{j}\left(\alpha\right)\underset{m=j+2}{\overset{m_{c}}{\Pi}}\left(1+\frac{\sigma_{m}\left(\alpha\right)}{\sigma_{m-1}\left(\alpha\right)}\right) & =\sigma_{j}\left(\alpha\right)\underset{m=2}{\overset{m_{c}-j}{\Pi}}\left(1+\frac{\sigma_{j+m}\left(\alpha\right)}{\sigma_{j+m-1}\left(\alpha\right)}\right)=\nonumber \\
 & =\sigma_{j}\left(\alpha\right)\underset{m=2}{\overset{m_{c}}{\Pi}}\left(1+\frac{\sigma_{j+m}\left(\alpha\right)}{\sigma_{j+m-1}\left(\alpha\right)}\right)+o\left(\mathcal{L}^{-1}\right)=\nonumber \\
 & =\sigma_{\mathrm{right}}\left(\alpha+2\pi j\right)+o\left(\mathcal{L}^{-1}\right).\label{eq: non-negative j's}
\end{align}
It is also evident from the relations in (\ref{eq: 2 pi shift relations})
that for $m_{1},m_{2}\ge1$ such that $m_{1}+m_{2}>m_{c}$, 
\begin{equation}
\frac{\sigma_{-m_{1}}\left(\alpha\right)}{\sigma_{-m_{1}-1}\left(\alpha\right)}\cdot\frac{\sigma_{m_{2}}\left(\alpha\right)}{\sigma_{m_{2}-1}\left(\alpha\right)}=o\left(\mathcal{L}^{-1}\right).
\end{equation}
We can thus conclude that for every $-m_{c}\le j<0$,
\begin{align}
\sigma_{j}\left(\alpha\right)\underset{m=j+2}{\overset{m_{c}}{\Pi}}\left(1+\frac{\sigma_{m}\left(\alpha\right)}{\sigma_{m-1}\left(\alpha\right)}\right) & =\sigma_{j}\left(\alpha\right)\underset{m=j+2}{\overset{m_{c}+j}{\Pi}}\left(1+\frac{\sigma_{m}\left(\alpha\right)}{\sigma_{m-1}\left(\alpha\right)}\right)+o\left(\mathcal{L}^{-1}\right)=\nonumber \\
 & =\sigma_{\mathrm{right}}\left(\alpha+2\pi j\right)+o\left(\mathcal{L}^{-1}\right).\label{eq: negative j's}
\end{align}
From (\ref{eq: Renyi rewritten}), (\ref{eq: non-negative j's})
and (\ref{eq: negative j's}) we can now derive that
\begin{equation}
S_{n}\left(\alpha\right)=\underset{j=-m_{c}}{\overset{m_{c}}{\sum}}\sigma_{\mathrm{right}}\left(\alpha+2\pi j\right)+o\left(\mathcal{L}^{-1}\right).
\end{equation}

The symmetry of the expression for $S_{n}\left(\alpha\right)$ in
(\ref{eq: Renyi truncated multiple}) obviously enables us to equivalently
write
\begin{equation}
S_{n}\left(\alpha\right)=\underset{j=-m_{c}}{\overset{m_{c}}{\sum}}\sigma_{\mathrm{left}}\left(\alpha+2\pi j\right)+o\left(\mathcal{L}^{-1}\right),
\end{equation}
where we have defined
\begin{equation}
\sigma_{\mathrm{left}}\left(\alpha\right)\equiv\sigma_{0}\left(\alpha\right)\left[\underset{m=2}{\overset{m_{c}}{\Pi}}\left(1+\frac{\sigma_{-m}\left(\alpha\right)}{\sigma_{-m+1}\left(\alpha\right)}\right)\right].
\end{equation}
This means that we can define 
\begin{equation}
\tilde{S}_{n}\left(\alpha\right)\equiv\frac{\sigma_{\mathrm{left}}\left(\alpha\right)+\sigma_{\mathrm{right}}\left(\alpha\right)}{2}=\frac{\sigma_{0}\left(\alpha\right)}{2}\left[\underset{m=2}{\overset{m_{c}}{\Pi}}\left(1+\frac{\sigma_{-m}\left(\alpha\right)}{\sigma_{-m+1}\left(\alpha\right)}\right)+\underset{m=2}{\overset{m_{c}}{\Pi}}\left(1+\frac{\sigma_{m}\left(\alpha\right)}{\sigma_{m-1}\left(\alpha\right)}\right)\right],
\end{equation}
and obtain the desired structure, namely 
\begin{equation}
S_{n}\left(\alpha\right)=\underset{j=-m_{c}}{\overset{m_{c}}{\sum}}\tilde{S}_{n}\left(\alpha+2\pi j\right)+o\left(\mathcal{L}^{-1}\right).
\end{equation}

\section{\label{sec:Symmetry-resolved-entanglement XY}Symmetry-resolved EE
for the XY model}

We now derive the asymptotic behavior of the analog of the flux-resolved
REE for the ground state of the XY model, namely the parity-resolved
$S_{n}^{\left(-\right)}\equiv\text{Tr}\left(\rho_{A}^{n}\left(-1\right)^{\hat{Q}_{A}}\right)$.
We assume for simplicity that $h\ge0$. Using (\ref{eq: RDM using fermionic operators}),
we can write
\begin{equation}
S_{n}^{\left(\pm\right)}=\underset{m=1}{\overset{L}{\Pi}}\left[\left(\frac{1-\nu_{m}}{2}\right)^{n}\pm\left(\frac{1+\nu_{m}}{2}\right)^{n}\right],\label{eq: XY REE explicit}
\end{equation}
where we also denoted $S_{n}^{\left(+\right)}\equiv S_{n}$. Note
that in particular we can immediately deduce that $S_{1}^{\left(-\right)}=S_{2}^{\left(-\right)}$.

\subsection{Gapped XY model}

We first estimate $S_{n}^{\left(-\right)}$ at the
limit $L\rightarrow\infty$ assuming that the system is gapped, i.e.,
$h\neq2$. As was explained in \ref{subsec:XY-model}, as $L\rightarrow\infty$
the values $\pm\nu_{m}$ converge in pairs to the values $\lambda_{l}$
defined in (\ref{eq: Limit of nu_m}), which in turn depend on $h$.

The case $h<2$ is simple: since $\lambda_{0}=0$, we obtain $S_{n}^{\left(-\right)}\rightarrow0$.
For $h>2$, on the other hand, the asymptotic expression for $S_{n}^{\left(-\right)}$
does not vanish. Indeed, we can write
\begin{align}
\underset{L\rightarrow\infty}{\lim}\underset{m=1}{\overset{L}{\Pi}}\left|\left(\frac{1-\nu_{m}}{2}\right)^{n}-\left(\frac{1+\nu_{m}}{2}\right)^{n}\right|= & \underset{m=-\infty}{\overset{\infty}{\Pi}}\left|\left(\frac{1-\lambda_{m}}{2}\right)^{n}-\left(\frac{1+\lambda_{m}}{2}\right)^{n}\right|,\\
\nonumber 
\end{align}
and writing $q\equiv e^{-\pi\tau_{0}}$ ($\tau_{0}$ was defined in
(\ref{eq: definition of tau_0})) we get
\begin{equation}
\underset{m=-\infty}{\overset{\infty}{\Pi}}\left|\left(\frac{1-\lambda_{m}}{2}\right)^{n}-\left(\frac{1+\lambda_{m}}{2}\right)^{n}\right|=\underset{m=0}{\overset{\infty}{\Pi}}\left[\left(\frac{1}{1+q^{2m+1}}\right)^{n}-\left(\frac{q^{2m+1}}{1+q^{2m+1}}\right)^{n}\right]^{2},
\end{equation}
so that eventually we obtain 
\begin{equation}
\underset{L\rightarrow\infty}{\lim}\left|S_{n}^{\left(-\right)}\right|=\underset{m=0}{\overset{\infty}{\Pi}}\left[\left(\frac{1}{1+q^{2m+1}}\right)^{n}-\left(\frac{q^{2m+1}}{1+q^{2m+1}}\right)^{n}\right]^{2}=\left(\frac{\underset{m=0}{\overset{\infty}{\Pi}}\left[1-q^{n\left(2m+1\right)}\right]}{\underset{m=0}{\overset{\infty}{\Pi}}\left[1+q^{\left(2m+1\right)}\right]^{n}}\right)^{2}.
\end{equation}

In order to further simplify this result for $\underset{L\rightarrow\infty}{\lim}\left|S_{n}^{\left(-\right)}\right|$,
we remind the reader of the definition of the Jacobi theta functions
\cite{whittaker_watson_1996}:
\begin{align}
\vartheta_{2}\left(z,q\right) & =\underset{m=-\infty}{\overset{\infty}{\sum}}q^{\left(m+\frac{1}{2}\right)^{2}}e^{2iz\left(m+\frac{1}{2}\right)},\nonumber \\
\vartheta_{3}\left(z,q\right) & =\underset{m=-\infty}{\overset{\infty}{\sum}}q^{m^{2}}e^{2izm},\nonumber \\
\vartheta_{4}\left(z,q\right) & =\underset{m=-\infty}{\overset{\infty}{\sum}}\left(-1\right)^{m}q^{m^{2}}e^{2izm}.
\end{align}
We write $\theta_{j}\left(q\right)\equiv\vartheta_{j}\left(0,q\right)$
and 
\begin{equation}
k\left(q\right)\equiv\frac{\theta_{2}^{2}\left(q\right)}{\theta_{3}^{2}\left(q\right)},\,\,\,k'\left(q\right)\equiv\sqrt{1-k^{2}\left(q\right)}=\frac{\theta_{4}^{2}\left(q\right)}{\theta_{3}^{2}\left(q\right)}.
\end{equation}
This definition of $k$ implies that $q=\exp\left[-\pi\frac{I\left(k'\right)}{I\left(k\right)}\right]$
($I$ was defined in (\ref{eq: First elliptic integral})) \cite{whittaker_watson_1996},
and thus it agrees with the definition of $k$ previously presented
in (\ref{eq: Definition of k}). We also write $k_{n}\left(q\right)\equiv k\left(q^{n}\right)$
and $k_{n}'\left(q\right)\equiv k'\left(q^{n}\right)$, and rely on
the following identities from \cite{whittaker_watson_1996} that hold
for every $0<q<1$:
\begin{align}
\underset{m=0}{\overset{\infty}{\Pi}}\left[1+q^{\left(2m+1\right)}\right] & =\left(\frac{16q}{k^{2}k'^{2}}\right)^{\frac{1}{24}},\nonumber \\
\underset{m=0}{\overset{\infty}{\Pi}}\left[1-q^{\left(2m+1\right)}\right] & =k'^{\frac{1}{4}}\underset{m=0}{\overset{\infty}{\Pi}}\left[1+q^{\left(2m+1\right)}\right]=\left(\frac{16qk'^{4}}{k^{2}}\right)^{\frac{1}{24}}.
\end{align}
We then obtain that for $h>2$,
\begin{equation}
\underset{L\rightarrow\infty}{\lim}\left|S_{n}^{\left(-\right)}\right|=\left[\frac{\left(kk'\right)^{2n}k_{n}'^{4}}{16^{n-1}k_{n}^{2}}\right]^{\frac{1}{12}}.
\end{equation}

Since $S_{n}^{\left(-\right)}$ is real by definition we can only
have $S_{n}^{\left(-\right)}=\pm\left|S_{n}^{\left(-\right)}\right|$,
but this still leaves us with an ambiguity regarding the sign of $S_{n}^{\left(-\right)}$.
To resolve this ambiguity we turn to the large $h$ limit of the above
expression. The definition of $k$ in (\ref{eq: Definition of k})
implies that as $h\rightarrow\infty$, $k\rightarrow0$ and therefore
$k'\rightarrow1$ and $q\rightarrow0$. Furthermore, one can show
that as $q\rightarrow0$, $k\sim4q^{\frac{1}{2}}$ \cite{whittaker_watson_1996}
and consequently
\begin{equation}
\underset{h\rightarrow\infty}{\lim}\left[\frac{\left(kk'\right)^{2n}k_{n}'^{4}}{16^{n-1}k_{n}^{2}}\right]^{\frac{1}{12}}=1.
\end{equation}
On the other hand, as can be easily seen from the Hamiltonian in (\ref{eq: JW Hamiltonian}),
in the large $h$ limit the system in question is ferromagnetic, and
we therefore expect that as $h\rightarrow\infty$ all $L$ fermion
sites of subsystem $A$ will be occupied in the ground state (i.e.,
$\rho_{A}$ has a non-vanishing eigenvalue only for the state that
corresponds to $Q_{A}=L$). This, in turn, suggests that for every
finite $L$, as $h\rightarrow\infty$ we obtain $S_{n}^{\left(-\right)}\rightarrow1$
for even $L$ and $S_{n}^{\left(-\right)}\rightarrow-1$ for odd $L$.
By continuity, the sign should remain the same for finite $h>2$.

This finally brings us to 
\begin{equation}
\underset{L\rightarrow\infty}{\lim}\left(-1\right)^{L}S_{n}^{\left(-\right)}=\begin{cases}
0, & h<2\\
\left[\frac{\left(kk'\right)^{2n}k_{n}'^{4}}{16^{n-1}k_{n}^{2}}\right]^{\frac{1}{12}}, & h>2
\end{cases}.\label{eq: XY Renyi symmetry block}
\end{equation}

\begin{figure}[h]
\begin{centering}
\includegraphics[scale=0.5]{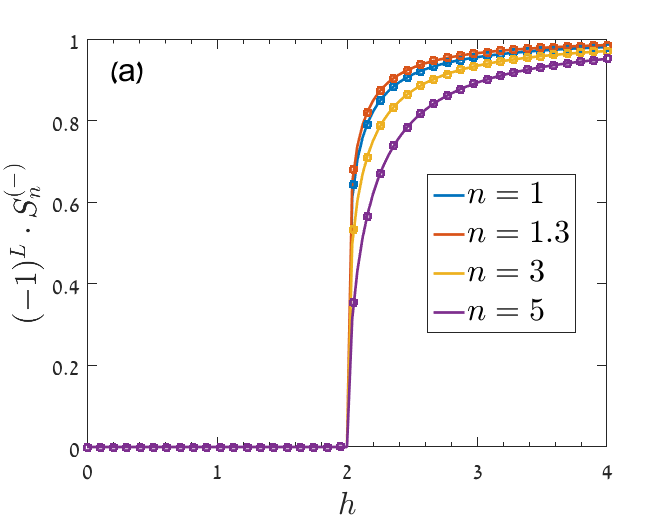}\includegraphics[scale=0.5]{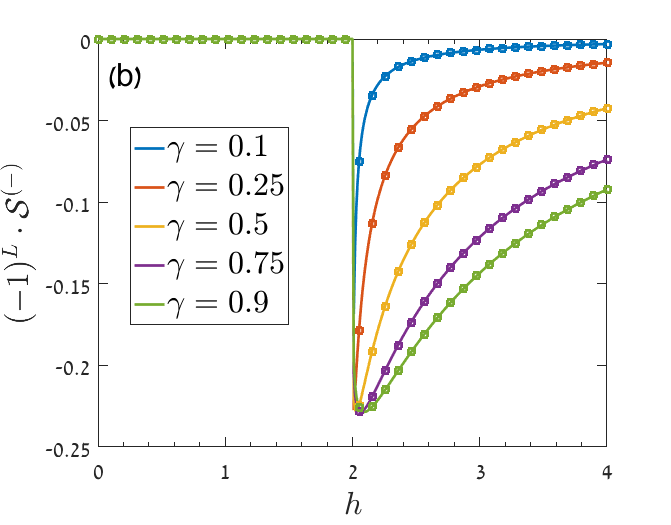}
\par\end{centering}
\caption{\label{fig: Renyi XY}(a) $\left(-1\right)^{L}S_{n}^{\left(-\right)}$
in a subsystem of $L$ sites of a gapped XY chain, for anisotropy
factor $\gamma=0.5$. The results were computed numerically for $L=200$
using (\ref{eq: XY REE explicit}) (dots) and analytically for $L\rightarrow\infty$
using (\ref{eq: XY Renyi symmetry block}) (continuous lines). (b)
$\left(-1\right)^{L}\mathcal{S}^{\left(-\right)}$ in a subsystem
of $L$ sites of a gapped XY chain, computed numerically for $L=200$
using (\ref{eq: XY vNEE explicit}) (dots) and analytically for $L\rightarrow\infty$
using (\ref{eq: XY vNEE}) (continuous lines).}
\end{figure}
Fig. \ref{fig: Renyi XY}(a) shows a comparison between the asymptotic
analytical result for $S_{n}^{\left(-\right)}$ and the numerical
result. It indicates a very good agreement between the two calculations,
and in particular confirms two conspicuous properties of the analytical
result in the large $L$ limit: that $S_{n}^{\left(-\right)}\rightarrow0$
in the $h<2$ regime, and that $\left|S_{n}^{\left(-\right)}\right|\rightarrow1$
as $h\rightarrow\infty$. A numerical calculation of $S_{2}^{\left(-\right)}$
for several values of $L$ has previously appeared in \cite{PhysRevB.99.115429}.

We use our calculation of $S_{n}^{\left(-\right)}$ in order to calculate
$\mathcal{S}^{\left(-\right)}=-\underset{n\rightarrow1}{\lim}\partial_{n}S_{n}^{\left(-\right)}$
at the large $L$ limit. Relying on (\ref{eq: XY REE explicit}),
we can obtain an explicit expression for $\mathcal{S}^{\left(-\right)}$:
\begin{equation}
\mathcal{S}^{\left(-\right)}=\left(-1\right)^{L}\underset{m=1}{\overset{L}{\sum}}\left(\underset{j\neq m}{\Pi}\nu_{j}\right)\cdot\left[\frac{1-\nu_{m}}{2}\ln\left(\frac{1-\nu_{m}}{2}\right)-\frac{1+\nu_{m}}{2}\ln\left(\frac{1+\nu_{m}}{2}\right)\right].\label{eq: XY vNEE explicit}
\end{equation}
This expression can be used for numerical estimates of ${\cal S}^{\left(-\right)}$.

\begin{figure}[h]
\begin{centering}
\includegraphics[scale=0.5]{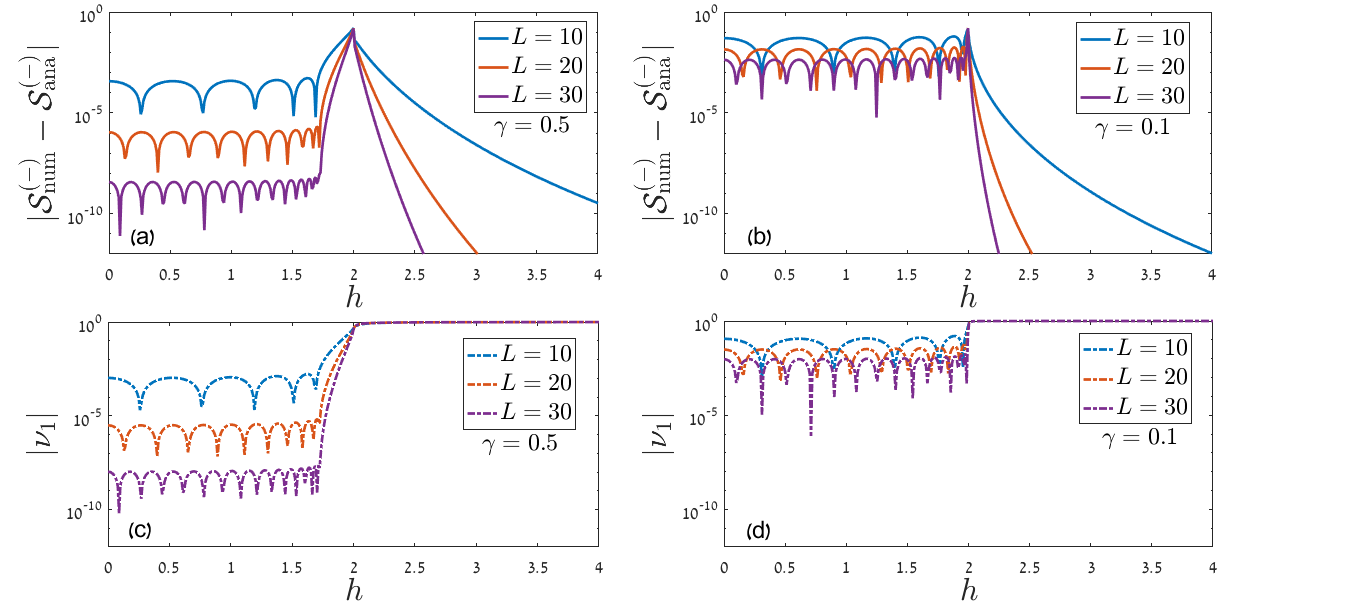}
\par\end{centering}
\caption{\label{fig: vNEE XY comparison}Upper panels: Absolute deviation of
the analytical estimate (\ref{eq: XY vNEE}) of $\mathcal{S}^{\left(-\right)}$
as $L\rightarrow\infty$ (denoted by $\mathcal{S}_{\mathrm{ana}}^{\left(-\right)}$)
from the numerical estimate (\ref{eq: XY vNEE explicit}) for finite
$L$ (denoted by $\mathcal{S}_{\mathrm{num}}^{\left(-\right)}$),
for (a) $\gamma=0.5$ and (b) $\gamma=0.1$. Lower panels: $\left|\nu_{1}\right|$
as a function of the magnetic field, for (c) $\gamma=0.5$ and (d)
$\gamma=0.1$. The minima that appear for $h<2$ in all of the graphs
correspond to points where $\left|\nu_{1}\right|$ vanishes and therefore
${\cal S}_{\mathrm{num}}^{\left(-\right)}$ vanishes as well.}

\end{figure}

From (\ref{eq: XY Renyi symmetry block}) we can now calculate $\mathcal{S}^{\left(-\right)}$
as $L\rightarrow\infty$: 
\begin{equation}
\underset{L\rightarrow\infty}{\lim}\left(-1\right)^{L}\mathcal{S}^{\left(-\right)}=\begin{cases}
0, & h<2\\
\frac{\sqrt{k'}}{3}\left[\ln2-\frac{1}{2}\ln\left(k\cdot k'\right)-\frac{I\left(k\right)I\left(k'\right)}{\pi}\left(1+k^{2}\right)\right], & h>2
\end{cases}.\label{eq: XY vNEE}
\end{equation}
The details of this calculation appear in subsection \ref{subsec: appendix 3}
of the appendix. ${\cal S}^{\left(-\right)}$ is plotted in Fig. \ref{fig: Renyi XY}(b),
where again good agreement between the analytical estimate and the
numerical result is evident. Figs. \ref{fig: vNEE XY comparison}(a)-(b)
show the difference between the analytical limit for $L\rightarrow\infty$
and the numerical results for finite $L$. They demonstrate that away
from the vicinity of $h=2$, where the phase transition occurs, corrections
to the asymptotic result vanish rapidly as $L$ grows, and it is apparent
that e.g. for $\gamma=0.5$ these corrections turn negligible even
for a relatively short subsystem. As $h$ nears $h=2$, we need a
larger value of $L$ in order for the deviation to be small.

Both $S_{n}^{\left(-\right)}$ and ${\cal S}^{\left(-\right)}$ illustrate
a striking property of the phase in which the system is found for
$h<2$: since $S_{n}^{\left(-\right)}={\cal S}^{\left(-\right)}=0$,
we obtain that for $h<2$, the system satisfies $S_{n}^{\left(\mathrm{even}\right)}=S_{n}^{\left(\mathrm{odd}\right)}$
and ${\cal S}^{\left(\mathrm{even}\right)}={\cal S}^{\left(\mathrm{odd}\right)}$.
This property stems from the fact that we can write the RDM as $\rho_{A}=\exp\left(-H_{A}\right)$
where the entanglement Hamiltonian $H_{A}$ is quadratic \cite{universe5010033,Peschel_2003},
and treat $H_{A}$ as the Hamiltonian of an effective system of a
1D open fermionic chain with $L$ sites. $H_{A}$ is expected to have
the same modes at the virtual edges of the subsystem as the original
system (the Kitaev chain) would host at a physical edge \cite{PhysRevLett.101.010504}.
Thus the phase $h<2$ corresponds to a topologically non-trivial phase
of $H_{A}$ where two Majorana zero-modes --- one at each end of
the system --- remain decoupled, provided that the virtual chain
is long enough \cite{Leijnse_2012}. Combining these two Majorana
operators yields a fermionic operator whose occupancy does not change
the eigenvalues of $H_{A}$, and thus induces a two-fold degeneracy
in the system: every eigenstate of $H_{A}$ with an even total fermionic
number has a corresponding eigenstate with the same eignevalue but
with an odd total fermionic number, and vice versa. This degeneracy
persists as long as $h<2$. This explains why in the large $L$ limit,
the contributions to the entropy from the block that corresponds to
an even $Q_{A}$ and the block that corresponds to an odd $Q_{A}$
are exactly the same. Our work provides a rigorous proof of this behavior
for the system considered. These observations allow us to explain
the finite $L$ corrections to our results, which become noticeable
for $L\lesssim50$, as depicted in Figs. \ref{fig: vNEE XY comparison}(a)-(b).

Since for $h\neq2$ the system is gapped, the correlations vanish
exponentially as $L\rightarrow\infty$ \cite{PhysRevB.69.104431},
and therefore so do the corrections to the limiting values of\textbf{
}$S_{n}^{\left(-\right)}$ and $\mathcal{S}^{\left(-\right)}$. For
$h<2$ the corrections are dominated by the hybridization of the entanglement
Majorana edge-modes: though they are localized exponentially at the
ends of the virtual chain \cite{Leijnse_2012}, for finite $L$ the
virtual edge Majorana fermions exhibit some overlap, and therefore
a true degeneracy is not achieved \cite{Leijnse_2012} for most values
of $h<2$, resulting in a finite nonzero value of the lowest eigenvalue
$\left|\nu_{1}\right|\equiv\min\left|\nu_{m}\right|$. Yet for certain
values of $h<2$ the virtual Majorana wave functions interfere destructively,
and this creates the minima apparent in Figs. \ref{fig: vNEE XY comparison}(a)-(d)
in both $\left|\mathcal{S}^{\left(-\right)}\right|$ and $\left|\nu_{1}\right|$.
This in fact suggests that the finite size corrections to $\mathcal{S}^{\left(-\right)}$
are dominated by $\nu_{1}$,
\begin{equation}
\left|{\cal S}^{\left(-\right)}\right|\approx\left|\frac{1-\nu_{1}}{2}\ln\left(\frac{1-\nu_{1}}{2}\right)-\frac{1+\nu_{1}}{2}\ln\left(\frac{1+\nu_{1}}{2}\right)\right|.
\end{equation}
The accuracy of this relation can serve as a quantitative test for
the above arguments. And indeed, calculating the ratio between this
approximation and $\left|{\cal S}^{\left(-\right)}\right|$ for the
cases that appear in Figs. \ref{fig: vNEE XY comparison}(a)-(b),
we get that for $h<1.9$ (outside the vicinity of the critical point
$h=2$), the contribution of $\nu_{1}$ to ${\cal S}^{\left(-\right)}$
is always above $85\%$ for $\gamma=0.5$, and always above $65\%$
for $\gamma=0.1$.

Considerations similar to those detailed above allow the extension
of our main results (\ref{eq: XY Renyi symmetry block}) and (\ref{eq: XY vNEE})
to $0>h\neq-2$. The limit $\underset{L\rightarrow\infty}{\lim}\left|S_{n}^{\left(-\right)}\right|$
is symmetric in $h$, and therefore, in particular, it tends to $1$
as $h\rightarrow-\infty$. However, the sign ambiguity is resolved
in a different way than in the $h>0$ case: for finite $L$ we expect
that in the $h\rightarrow-\infty$ limit, all sites of $A$ become
unoccupied such that $Q_{A}=0$ with probability $1$. We thus obtain
that as $h\rightarrow-\infty$, $S_{n}^{\left(-\right)}\rightarrow1$
both for even and odd $L$, and so for $0>h\neq-2$ the limit $\underset{L\rightarrow\infty}{\lim}S_{n}^{\left(-\right)}$
exists and is also positive.

The extensions of (\ref{eq: XY Renyi symmetry block}) and (\ref{eq: XY vNEE})
to $0>h\neq-2$ are therefore symmetric, apart from the absence of
the $\left(-1\right)^{L}$ prefix, namely 
\begin{equation}
\underset{L\rightarrow\infty}{\lim}S_{n}^{\left(-\right)}\Big\rvert_{h}=\underset{L\rightarrow\infty}{\lim}\left(-1\right)^{L}S_{n}^{\left(-\right)}\Big\rvert_{-h}\,\,\,\left(0>h\neq-2\right),
\end{equation}
and
\begin{equation}
\underset{L\rightarrow\infty}{\lim}{\cal S}^{\left(-\right)}\Big\rvert_{h}=\underset{L\rightarrow\infty}{\lim}\left(-1\right)^{L}{\cal S}^{\left(-\right)}\Big\rvert_{-h}\,\,\,\left(0>h\neq-2\right).
\end{equation}

\subsection{Gapless XY model}

Here we calculate the asymptotics of $S_{n}^{\left(\pm\right)}$
for the case where the system is gapless, i.e., $h=2$. Following
\cite{Its_2005}{} we write
\begin{equation}
\ln\left|S_{n}^{\left(\pm\right)}\right|=\mathrm{Re}\left\{ \underset{\varepsilon,\delta\rightarrow0^{+}}{\lim}\frac{1}{4\pi i}\underset{c\left(\varepsilon,\delta\right)}{\int}e_{n}^{\left(\pm\right)}\left(1+\varepsilon,\lambda\right)\frac{d}{d\lambda}\ln\tilde{D}_{L}\left(\lambda\right)d\lambda\right\} ,
\end{equation}
where $c\left(\varepsilon,\delta\right)$ is the contour shown in
Fig. \ref{fig:The-integration-contour}(a), and
\begin{equation}
e_{n}^{\left(\pm\right)}\left(x,\nu\right)\equiv\ln\left[\left(\frac{x-\nu}{2}\right)^{n}\pm\left(\frac{x+\nu}{2}\right)^{n}\right].
\end{equation}
Using the asymptotic approximation for $\tilde{D}_{L}\left(\lambda\right)$
in (\ref{eq: Asymptotic det gapless XY}), we obtain that
\begin{equation}
\frac{d}{d\lambda}\ln\tilde{D}_{L}\left(\lambda\right)\sim\left(\frac{1}{\lambda+1}+\frac{1}{\lambda-1}\right)L-\frac{4i}{\pi}\cdot\frac{\beta\left(\lambda\right)}{\left(\lambda+1\right)\left(\lambda-1\right)}\ln L.
\end{equation}
This expression is reminiscent of (\ref{eq: Fisher Hartwig dlog})
from the calculation of the REE in the gapless XX model, and we can
therefore carry out the integration along the same lines of argument,
so that eventually we get
\begin{equation}
\ln\left|S_{n}^{\left(\pm\right)}\right|\sim\left[\frac{1}{12}\left(\frac{1}{n}-n\right)-\frac{\eta_{\pm}}{4n}\right]\ln L,
\end{equation}
where $\eta_{+}=0$ and $\eta_{-}=1$. Since $S_{n}^{\left(+\right)}=\left|S_{n}^{\left(+\right)}\right|$,
we have in particular obtained that for the gapless XY model
\begin{equation}
\ln S_{n}^{\left(+\right)}\sim\frac{1}{12}\left(\frac{1}{n}-n\right)\ln L,
\end{equation}
a special case of a result which was derived and verified numerically
in \cite{PhysRevA.92.042334}. The coefficient of the logarithm is halved
as compared to (\ref{eq: 0th order expression}) with $\alpha=0$,
since a Majorana mode rather than a complex fermion is gapless here,
in accordance with CFT predictions \cite{2009JPhA...42X4005C}.

\begin{figure}[h]
\begin{centering}
\includegraphics[scale=0.5]{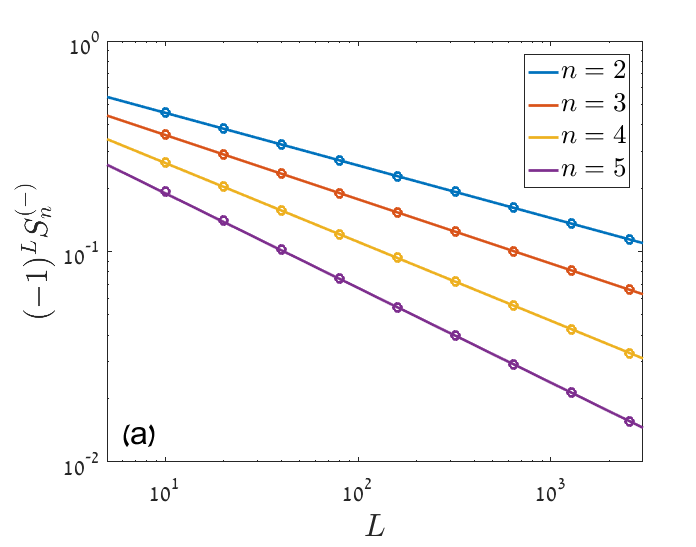}\includegraphics[scale=0.5]{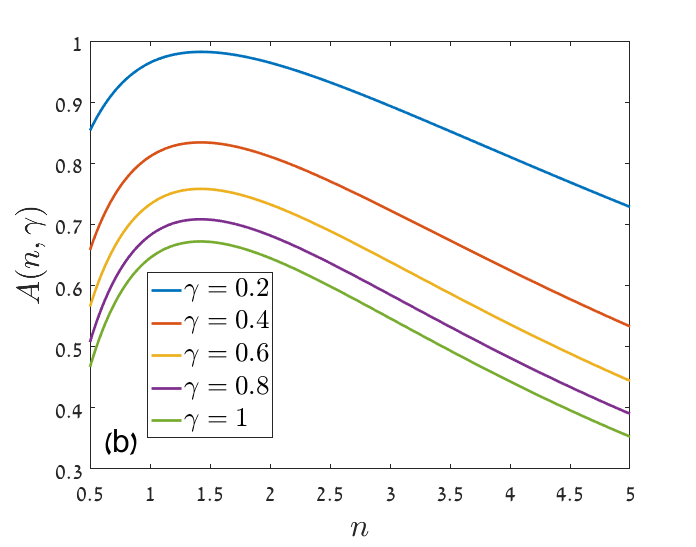}
\par\end{centering}
\caption{\label{fig: Gapless XY}(a) $\left(-1\right)^{L}S_{n}^{\left(-\right)}$
in a subsystem of $L$ sites of a gapless XY chain ($h=2)$, for anisotropy
factor $\gamma=0.4$. The results were computed numerically using
(\ref{eq: XY REE explicit}) (dots) and analytically using (\ref{eq: Ungapped XY parity-resolved Renyi})
(continuous lines). For the analytical results, the unknown factor
$A\left(n,\gamma\right)$ was extracted from a fit to the numerical
results. (b) The factor $A\left(n,\gamma\right)$ for several values
of anisotropy factor $\gamma$, as extracted from fits to numerical
results.}

\end{figure}

$S_{n}^{\left(-\right)}$ is again determined only
up to a sign,
\begin{equation}
\left|S_{n}^{\left(-\right)}\right|\approx A\left(n,\gamma\right)L^{-\frac{1}{6n}-\frac{n}{12}},
\end{equation}
where $A\left(n,\gamma\right)$ is some positive factor independent
of $L$, assuming that the largest subleading contribution not appearing
in the approximation (\ref{eq: Asymptotic det gapless XY}) does not
depend on $L$. We have verified this assumption numerically by fitting
to the results a function of $L$ that scales as $L^{-1/6n-n/12}$,
while the proportionality constant remained a free parameter. An example
is shown in Fig. \ref{fig: Gapless XY}(a), where good agreement between
numerical and analytical results is evident. Lead by similar considerations
as in the case of the gapped XY model, we can determine the sign of
$S_{n}^{\left(-\right)}$ to be $\left(-1\right)^{L}$, so that
\begin{equation}
\left(-1\right)^{L}S_{n}^{\left(-\right)}\approx A\left(n,\gamma\right)L^{-\frac{1}{6n}-\frac{n}{12}}.\label{eq: Ungapped XY parity-resolved Renyi}
\end{equation}
Specific attributes of the factor $A\left(n,\gamma\right)$ are generally
not captured by known theorems or conjectures we are aware of, and
its analysis is beyond the scope of this work. We show its typical
behavior, as extracted from numerical results, in Fig. \ref{fig: Gapless XY}(b).
The result (\ref{eq: Ungapped XY parity-resolved Renyi}) confirms
a previous prediction based on CFT considerations \cite{PhysRevLett.120.200602}.

The parity-resolved vNEE $\mathcal{S}^{\left(-\right)}$
for $h=2$ is therefore
\begin{equation}
\left(-1\right)^{L}\mathcal{S}^{\left(-\right)}\approx-\frac{A\left(1,\gamma\right)}{12}L^{-\frac{1}{4}}\ln L.
\end{equation}
As before, we can extend the results for $S_{n}^{\left(-\right)}$
and $\mathcal{S}^{\left(-\right)}$ to $h=-2$ by simply omitting
the $\left(-1\right)^{L}$ prefix.

\section{\label{sec:Generalization-to-higher}Generalization to higher dimensions}

In order to find the leading asymptotic behavior of the charged-resolved
REE in a $d$-dimensional gapless free Fermi gas, we rely in this
section on a formula conjectured by Widom \cite{widom1982class} and
proven for several particular cases \cite{sobolev2013pseudo,PhysRevLett.112.160403,Sobolev2015}.
A result similar to that which we are about to present was derived
in a recent work \cite{PhysRevLett.121.150501}, which discussed a
different but related quantity, the accessible EE defined there.

Let us describe the physical scale of subsystem $A$ in terms of a
typical linear dimension $L\gg1$ (made dimensionless by e.g. normalizing
by the lattice constant), so that $A$ contains $L^{d}$ sites. We
denote by $\Omega_{A}$ the bounded region in real space that is occupied
by $A$, and by $\Gamma$ the region in momentum space that is occupied
by the Fermi sea. We further denote by $P$ and $Q$ the operators
which represent projections into $\Gamma$ and $\Omega_{A}$, respectively.
Following \cite{PhysRevLett.120.200602}, we can write
\begin{equation}
\ln S_{n}\left(\alpha\right)=\mathrm{Tr}\ln\left[C^{n}e^{i\alpha}+\left(1-C\right)^{n}\right],
\end{equation}
where $C_{ij}=\langle a_{i}^{\dagger}a_{j}\rangle$ ($i,j=1,\ldots,L^{d}$)
is the fermionic correlation matrix, restricted to subsystem $A$.
In the ground state $C=QPQ$, and therefore $\ln S_{n}\left(\alpha\right)=\mathrm{Tr}f_{n,\alpha}\left(QPQ\right)$,
where $f_{n,\alpha}\left(t\right)=\ln\left[t^{n}e^{i\alpha}+\left(1-t\right)^{n}\right]$.

We now introduce the notations
\begin{equation}
c_{1}=\frac{1}{\left(2\pi\right)^{d}L^{d}}\underset{\Omega_{A}}{\int}\underset{\Gamma}{\int}\text{d}\boldsymbol{x}\text{d}\boldsymbol{p}\,\,\,\text{and}\,\,\,c_{2}=\frac{1}{\left(2\pi\right)^{d+1}L^{d-1}}\underset{\partial\Omega_{A}}{\int}\underset{\partial\Gamma}{\int}\left|\boldsymbol{n_{x}}\cdot\boldsymbol{n_{p}}\right|\text{d}S_{x}\text{d}S_{p},
\end{equation}
where $\boldsymbol{n_{x}},\boldsymbol{n_{p}}$ are unit vectors that
are normal to $\partial\Omega_{A},\partial\Gamma$, respectively,
and the units of $\boldsymbol{x}$ are set by the lattice constant such that the volume of a lattice unit cell equals unity. The normalization by powers of $L$ was chosen so
that the scaling with $L$ of the final results becomes apparent. Note that in previous works \cite{PhysRevLett.121.150501,PhysRevLett.112.160403,PhysRevLett.96.100503,klich2008scaling} this was achieved by a different convention of measuring $\boldsymbol{x}$ in units of $L$.
A function $f$ is said to obey the Widom formula \cite{PhysRevLett.112.160403,PhysRevLett.96.100503,klich2008scaling}
if for $L\gg1$,
\begin{equation}
\mathrm{Tr}f\left(QPQ\right)=c_{1}f\left(1\right)L^{d}+c_{2}U\left(f\right)L^{d-1}\ln L+o\left(L^{d-1}\ln L\right),\label{eq: Widom formula}
\end{equation}
where we have defined 
\begin{equation}
U\left(f\right)\equiv\underset{0}{\overset{1}{\int}}\frac{f\left(t\right)-tf\left(1\right)}{t\left(1-t\right)}dt.\label{eq: definition of I(f)}
\end{equation}
Note that the formula (\ref{eq: Widom formula}) was proven rigorously
only for regions $\Omega_{A}$, $\Gamma$ which satisfy certain regularity
conditions, detailed in \cite{PhysRevLett.112.160403}.

In \cite{PhysRevLett.112.160403} it was shown that $f:\mathbb{R}\rightarrow\mathbb{R}$
satisfies the Widom formula in two specific cases:
\begin{lyxlist}{00.00.0000}
\item [{\textbf{Case~(a)}}] $f$ is infinitely differentiable and $f\left(0\right)=0$.
\item [{\textbf{Case~(b)}}] $f$ is infinitely differentiable on $\mathbb{R}\setminus\left\{ 0,1\right\} $
and there exist real constants $K,\beta>0$ so that for every $t\in\left[0,1\right]$,
$\left|f\left(t\right)\right|\le Kt^{\beta}\left(1-t\right)^{\beta}$.
\end{lyxlist}
Let us define $F_{n,\alpha}\left(t\right)=f_{n,\alpha}\left(t\right)-f_{n,\alpha}\left(1\right)t$
for every $n>0$ and $-\pi<\alpha<\pi$. Both the real and the imaginary
parts of $F_{n,\alpha}$ satisfy the requirements of Case (b)\footnote{For $\alpha=0$ we should define $F_{n,0}$ such that $F_{n,0}\left(t\right)=f_{n,0}\left(t\right)$
for $t\in\left[0,1\right]$ and $F_{n,0}(t)=0$ for $t\notin\left[0,1\right]$,
as was done in \cite{PhysRevLett.112.160403}.} with $\beta=\min\left\{ n,1\right\} $, and we can therefore apply
the Widom formula (\ref{eq: Widom formula}) to $F_{n,\alpha}$. Using
the fact that $F_{n,\alpha}\left(1\right)=0$ and $U\left(F_{n,\alpha}\right)=U\left(f_{n,\alpha}\right)$,
we obtain that
\begin{equation}
\mathrm{Tr}F_{n,\alpha}\left(QPQ\right)=c_{2}U\left(f_{n,\alpha}\right)L^{d-1}\ln L+o\left(L^{d-1}\ln L\right).
\end{equation}
The LHS of the last equality can be written as $\mathrm{Tr}F_{n,\alpha}\left(QPQ\right)=\mathrm{Tr}f_{n,\alpha}\left(QPQ\right)-f_{n,\alpha}\left(1\right)\mathrm{Tr}g\left(QPQ\right)$,
where $g\left(t\right)=t$. $g$ obeys the Widom formula because it
fulfills the requirements of Case (a), so by applying the Widom formula
to $\mathrm{Tr}g\left(QPQ\right)$ as well we can thus conclude that
\begin{equation}
\mathrm{Tr}f_{n,\alpha}\left(QPQ\right)=\mathrm{Tr}F_{n,\alpha}\left(QPQ\right)+f_{n,\alpha}\left(1\right)\mathrm{Tr}g\left(QPQ\right)=c_{1}f_{n,\alpha}\left(1\right)L^{d}+c_{2}U\left(f_{n,\alpha}\right)L^{d-1}\ln L+o\left(L^{d-1}\ln L\right),
\end{equation}
which shows that $f_{n,\alpha}$ itself obeys the Widom formula.

Consequently, we have for every $-\pi<\alpha<\pi$
\begin{equation}
\ln S_{n}\left(\alpha\right)=ic_{1}L^{d}\alpha+c_{2}U\left(f_{n,\alpha}\right)L^{d-1}\ln L+o\left(L^{d-1}\ln L\right).
\end{equation}
Substituting $f_{n,\alpha}$ into (\ref{eq: definition of I(f)})
and using the change of variables $u=\ln\frac{t}{1-t}$, we get
\begin{equation}
U\left(f_{n,\alpha}\right)=\underset{-\infty}{\overset{\infty}{\int}}\left[\ln\left(\frac{1+e^{nu+i\alpha}}{\left(1+e^{u}\right)^{n}}\right)-\frac{i\alpha}{1+e^{-u}}\right]du=\frac{\pi^{2}}{6}\left(\frac{1}{n}-n\right)-\frac{\alpha^{2}}{2n}.
\end{equation}
We can therefore write
\begin{equation}
S_{n}\left(\alpha\right)\approx\exp\left[ic_{1}L^{d}\alpha-\frac{1}{2}\cdot\frac{c_{2}L^{d-1}\ln L}{n}\alpha^{2}+\frac{\pi^{2}}{6}\left(\frac{1}{n}-n\right)c_{2}L^{d-1}\ln L\right],
\end{equation}
and finally conclude from (\ref{eq: flux-resolved to charge-resolved})
that in $d$ dimensions, the charge-resolved REE satisfies
\begin{equation}
S_{n}\left(Q_{A}\right)\approx\sqrt{\frac{n}{2\pi c_{2}L^{d-1}\ln L}}\exp\left[\frac{-n\left(Q_{A}-c_{1}L^{d}\right)^{2}}{2c_{2}L^{d-1}\ln L}+\frac{\pi^{2}}{6}\left(\frac{1}{n}-n\right)c_{2}L^{d-1}\ln L\right].
\end{equation}
For $d=1$, $c_{1}L^{d}=\langle Q_{A}\rangle$ and $c_{2}L^{d-1}=1/\pi^{2}$,
and therefore
\begin{equation}
S_{n}\left(Q_{A}\right)\approx S_{n}\cdot\sqrt{\frac{\pi n}{2\ln L}}\exp\left[\frac{-n\pi^{2}\left(Q_{A}-\langle Q_{A}\rangle\right)^{2}}{2\ln L}\right]\,\,\,\left(d=1\right),
\end{equation}
which is in complete agreement with the approximation (\ref{eq: charge-resolved XX gaussian})
to leading order in $\ln L/n$.

\section{\label{sec:Conclusions-and-future}Conclusions and future outlook}

In this work we have obtained analytically the asymptotic behavior
of the flux-resolved REE in a 1D spin (fermion) chain, both for a
gapless XX (tight binding) chain and for the XY (Kitaev)
chain, as well as in higher dimensions. In 1D, these analytical results
have been shown in general to be in very good agreement with numerical
results, even for a subsystem of moderate length.

For the gapless XX model our results agree with previous CFT arguments,
and extend them beyond leading order in $\mathcal{L}$. While the
Gaussian approximation and the leading order approximation of $S_{n}\left(\alpha\right)$
deviate considerably from numerical results, the approximation that
includes all terms up to order ${\cal O}\left({\cal L}^{-1}\right)$
has been extremely accurate in the cases we have examined. We were
also able to provide a meaning to the corrections beyond the leading
order approximation, by showing that they arise from a periodic structure,
in line with CFT arguments. In higher dimensions, we derived an approximated
expression for the symmetry-resolved REE in a gapless gas of free
fermions. Under such an approximation the symmetry-resolved EE is
proportionate to a Gaussian distribution of the charge, akin to the
equipartition property noted in \cite{PhysRevB.98.041106}.

For the gapped XY model, our results provide a way to obtain analytical
expressions for the parity-resolved decomposition of both the REE
and the vNEE. These expressions are, on the face of it, limiting expressions
that apply to a subsystem of infinite length, but our calculations
have shown that they match the numerical results even for relatively
short subsystems, due to the exponential decay of the correlations.
We have also detected a topologically non-trivial phase in the virtual
chain described by the entanglement Hamiltonian, which explains why
for $|h|<2$ there is an equal contribution to the EE from states where
$Q_{A}$ is odd and states where $Q_{A}$ is even. At
the critical points, $h=\pm2$, we have found a power-law behavior matching
previous CFT predictions \cite{PhysRevLett.120.200602}.

The use of the generalized Fisher-Hartwig (or, in higher dimensions,
the Widom) conjecture was thus proven to be a powerful method for
producing accurate estimates of symmetry-resolved EE. This suggests
several prospects of future research, applying similar methods of
calculation to questions such as the symmetry-resolved EE in topological
systems \cite{PhysRevB.99.115429,PhysRevLett.96.110404,PhysRevLett.96.110405},
or in systems out of equilibrium, for example following a quench \cite{2019arXiv190510749F}.
Another possible direction of research is the study
of the symmetry-resolved EE of a bipartition into disconnected subsystems
\cite{2019arXiv190904035F}.\\

\textit{Note added: }When this work was close to completion a related
work appeared online \cite{2019arXiv190702084B} which employs Fisher-Hartwig
techniques to calculate the resolved entropy of the XX chain to order
$\mathcal{O}\left(\mathcal{L}^{0}\right)$. Our results go further
in (i) performing the XX calculations to order $\mathcal{O}\left(\mathcal{L}^{-1}\right)$,
which is especially important in the vicinity of $\alpha=\pm\pi$;
(ii) studying the XY (Kitaev) case; (iii) treating higher-dimensional
gapless fermionic systems.

\section*{Acknowledgements}

We would like to thank P. Calabrese, P. Ruggiero and E. Sela for useful
discussions, and an anonymous referee who brought
to our attention the generalized Fisher-Hartwig conjecture of \cite{PhysRevA.92.042334,PhysRevA.97.062301}, which allowed us to treat the gapless XY case.
Support by the Israel Science Foundation (Grant No. 227/15) and the
US-Israel Binational Science Foundation (Grant No. 2016224) is gratefully
acknowledged.

\newpage{}

\appendix

\section{Appendix: Details of the calculations}

\setcounter{equation}{0}
\renewcommand{\theequation}{A.\arabic{equation}}

\subsection{Asymptotics of the correlation matrix determinant
(gapless XY model)\label{subsec: Appendix 4}}

We derive here the leading order asymptotic approximation
for the determinant $\tilde{D}_{L}\left(\lambda\right)$ defined in
(\ref{eq: definition of XY determinant}), for the case of a gapless
XY chain ($h=2$). A generalized version of the Fisher-Hartwig formula
was conjectured in \cite{PhysRevA.92.042334,PhysRevA.97.062301}, regarding the determinant
of a block Toeplitz matrix of the form
\begin{equation}
T_{L}\left[\mathcal{M}\right]=\left(\begin{array}{cccc}
\tilde{\Pi}_{0} & \tilde{\Pi}_{-1} & \cdots & \tilde{\Pi}_{1-L}\\
\tilde{\Pi}_{1} & \tilde{\Pi}_{0} &  & \vdots\\
\vdots &  & \ddots & \vdots\\
\tilde{\Pi}_{L-1} & \cdots & \cdots & \tilde{\Pi}_{0}
\end{array}\right),\,\,\,\tilde{\Pi}_{m}\equiv\frac{1}{2\pi}\underset{0}{\overset{2\pi}{\int}}d\theta e^{-im\theta}\mathcal{M}\left(\theta\right),
\end{equation}
where $\mathcal{M}\left(\theta\right)$ is a piecewise continuous
$d\times d$ matrix with jump discontinuities at the points $\theta_{r}$,
$r=0,\ldots,R$. We define for each discontinuity $\mathcal{M}_{r}^{\pm}\equiv\underset{\theta\rightarrow\theta_{r}^{\pm}}{\lim}\mathcal{M}\left(\theta\right)$,
and assume that for each $r$, $\mathcal{M}_{r}^{+}$ and $\mathcal{M}_{r}^{-}$
commute. This allows us to find a joint diagonalizing basis for $\mathcal{M}_{r}^{\pm}$,
and we denote the corresponding eigenvalues by $\mu_{r,j}^{\pm}$,
$j=1,\ldots,d$. According to the conjecture \cite{PhysRevA.97.062301},
for the first two leading terms of the large $L$ approximation of
$\ln\det T_{L}\left[\mathcal{M}\right]$ we then have
\begin{equation}
\ln\det T_{L}\left[\mathcal{M}\right]=\frac{L}{2\pi}\underset{0}{\overset{2\pi}{\int}}d\theta\ln\left(\det\mathcal{M}\left(\theta\right)\right)+\frac{\ln L}{4\pi^{2}}\underset{r=0}{\overset{R}{\sum}}\underset{j=1}{\overset{d}{\sum}}\left(\ln\left(\frac{\mu_{r,j}^{-}}{\mu_{r,j}^{+}}\right)\right)^{2}+\cdots.
\end{equation}

For $h=2$, $\tilde{D}_{L}\left(\lambda\right)$ is
of the form described above, i.e., $\tilde{D}_{L}\left(\lambda\right)=\det T_{L}\left[\mathcal{M}\right]$
for $\mathcal{M}\left(\theta\right)=i\lambda I_{2}-\mathcal{G}\left(\theta\right)$.
Now $\mathcal{M}$ has a single discontinuity at $\theta_{0}=0$,
with
\begin{equation}
\mathcal{M}_{0}^{\pm}=\left(\begin{array}{cc}
i\lambda & \pm i\\
\pm i & i\lambda
\end{array}\right),
\end{equation}
from which we obtain that $\mu_{0,1}^{\pm}=i\lambda\pm i$ and $\mu_{0,2}^{\pm}=i\lambda\mp i$.
Since $\det\mathcal{M}\left(\theta\right)=1-\lambda^{2}$ is independent
of $\theta$, we finally arrive at 
\begin{equation}
\ln\tilde{D}_{L}\left(\lambda\right)=\ln\left(1-\lambda^{2}\right)L-2\beta^{2}\left(\lambda\right)\ln L+\cdots.
\end{equation}

\subsection{Leading order approximation of the flux-resolved REE (XX model)\label{subsec: appendix 1}}

From the Fisher-Hartwig conjecture we have derived the leading order
approximation for the asymptotic expression for $S_{n}\left(\alpha\right)$:
\begin{align}
\ln S_{n}^{\left(0\right)}\left(\alpha\right) & =i\frac{\alpha}{2}L+\underset{\varepsilon,\delta\rightarrow0^{+}}{\lim}\frac{1}{2\pi i}\underset{c\left(\varepsilon,\delta\right)}{\int}e_{n}^{\left(\alpha\right)}\left(1+\varepsilon,\lambda\right)\left(\frac{k_{F}/\pi}{\lambda-1}+\frac{1-k_{F}/\pi}{\lambda+1}\right)Ld\lambda+\nonumber \\
 & +\underset{\varepsilon,\delta\rightarrow0^{+}}{\lim}\frac{1}{2\pi i}\underset{c\left(\varepsilon,\delta\right)}{\int}e_{n}^{\left(\alpha\right)}\left(1+\varepsilon,\lambda\right)\left(-\frac{4i}{\pi}\cdot\frac{\beta\left(\lambda\right)}{\left(\lambda+1\right)\left(\lambda-1\right)}\left[\ln\mathcal{L}+\left(1+\gamma_{E}\right)+\Upsilon\left(\lambda\right)\right]\right)d\lambda.
\end{align}
Regarding the first integral, it is easily shown that
\begin{equation}
\underset{\varepsilon,\delta\rightarrow0^{+}}{\lim}\frac{1}{2\pi i}\underset{c\left(\varepsilon,\delta\right)}{\int}e_{n}^{\left(\alpha\right)}\left(1+\varepsilon,\lambda\right)\left(\frac{k_{F}/\pi}{\lambda-1}+\frac{1-k_{F}/\pi}{\lambda+1}\right)Ld\lambda=i\left(-1+\frac{2k_{F}}{\pi}\right)\frac{\alpha}{2}L.
\end{equation}
As for the second integral, we use the fact that for every $-1<x<1$,
\begin{equation}
\beta\left(x+i0^{\pm}\right)=-iW\left(x\right)\mp\frac{1}{2},
\end{equation}
where $W\left(x\right)\equiv\frac{1}{2\pi}\ln\frac{1+x}{1-x}$. It
can be shown that the contribution from the circular arcs of the contour
$c\left(\varepsilon,\delta\right)$ vanishes as $\varepsilon,\delta\rightarrow0^{+},$
and therefore we get
\begin{align}
 & \underset{\varepsilon,\delta\rightarrow0^{+}}{\lim}\frac{1}{2\pi i}\underset{c\left(\varepsilon,\delta\right)}{\int}e_{n}^{\left(\alpha\right)}\left(1+\varepsilon,\lambda\right)\left(-\frac{4i}{\pi}\cdot\frac{\beta\left(\lambda\right)}{\left(\lambda+1\right)\left(\lambda-1\right)}\left[\ln\mathcal{L}+\left(1+\gamma_{E}\right)+\Upsilon\left(\lambda\right)\right]\right)d\lambda=\nonumber \\
 & =\frac{\left(\ln\mathcal{L}+\left(1+\gamma_{E}\right)\right)}{\pi^{2}}\underset{\varepsilon\rightarrow0^{+}}{\lim}\underset{-1+\frac{\varepsilon}{2}}{\overset{1-\frac{\varepsilon}{2}}{\int}}\frac{2e_{n}^{\left(\alpha\right)}\left(1+\varepsilon,x\right)}{1-x^{2}}dx+\nonumber \\
 & +\underset{\varepsilon\rightarrow0^{+}}{\lim}\underset{k=1}{\overset{\infty}{\sum}}\frac{1}{\pi^{2}k}\underset{-1+\frac{\varepsilon}{2}}{\overset{1-\frac{\varepsilon}{2}}{\int}}\left[\frac{\left(\frac{1}{2}-iW\left(x\right)\right)^{3}}{k^{2}-\left(\frac{1}{2}-iW\left(x\right)\right)^{2}}+\frac{\left(\frac{1}{2}+iW\left(x\right)\right)^{3}}{k^{2}-\left(\frac{1}{2}+iW\left(x\right)\right)^{2}}\right]\frac{2e_{n}^{\left(\alpha\right)}\left(1+\varepsilon,x\right)}{1-x^{2}}dx=\nonumber \\
 & =\frac{\ln\mathcal{L}}{\pi^{2}}\underset{\varepsilon\rightarrow0^{+}}{\lim}\underset{-1+\frac{\varepsilon}{2}}{\overset{1-\frac{\varepsilon}{2}}{\int}}\frac{2e_{n}^{\left(\alpha\right)}\left(1+\varepsilon,x\right)}{1-x^{2}}dx-\frac{1}{\pi^{2}}\underset{\varepsilon\rightarrow0^{+}}{\lim}\underset{-1+\frac{\varepsilon}{2}}{\overset{1-\frac{\varepsilon}{2}}{\int}}\left[\psi\left(\frac{1}{2}+iW\left(x\right)\right)+\psi\left(\frac{1}{2}-iW\left(x\right)\right)\right]\frac{e_{n}^{\left(\alpha\right)}\left(1+\varepsilon,x\right)}{1-x^{2}}dx.
\end{align}
Here we denoted by $\psi\left(x\right)$ the Digamma function, $\psi\left(x\right)=\frac{\Gamma'\left(x\right)}{\Gamma\left(x\right)}$,
and used the identity \cite{2004JSP...116...79J} 
\begin{equation}
\underset{k=1}{\overset{\infty}{\sum}}\frac{1}{k}\left[\frac{\left(\frac{1}{2}+iw\right)^{3}}{k^{2}-\left(\frac{1}{2}+iw\right)^{2}}+\frac{\left(\frac{1}{2}-iw\right)^{3}}{k^{2}-\left(\frac{1}{2}-iw\right)^{2}}\right]=-1-\gamma_{E}-\frac{1}{2}\left[\psi\left(\frac{1}{2}+iw\right)+\psi\left(\frac{1}{2}-iw\right)\right].
\end{equation}
Using a change of variables $u=\ln\frac{1+x}{1-x}$, and taking the
limit $\varepsilon\rightarrow0^{+}$, we have
\begin{align}
\underset{\varepsilon\rightarrow0^{+}}{\lim}\underset{-1+\frac{\varepsilon}{2}}{\overset{1-\frac{\varepsilon}{2}}{\int}}\frac{2e_{n}^{\left(\alpha\right)}\left(1+\varepsilon,x\right)}{1-x^{2}}dx & =-n\underset{-\infty}{\overset{\infty}{\int}}\frac{u}{e^{u}+1}\cdot\frac{e^{nu+i\frac{\alpha}{2}}-e^{u-i\frac{\alpha}{2}}}{e^{nu+i\frac{\alpha}{2}}+e^{-i\frac{\alpha}{2}}}du=\nonumber \\
 & =-\frac{\pi^{2}}{6}n+\frac{2}{n}\underset{k=1}{\overset{\infty}{\sum}}\frac{\left(-1\right)^{k+1}}{k^{2}}\cos\left(\alpha k\right).\\
\nonumber 
\end{align}
Recalling the Fourier series of $\alpha^{2}$ in $\left(-\pi,\pi\right)$,
we can now write for every $-\pi<\alpha<\pi$
\begin{equation}
\frac{\ln\mathcal{L}}{\pi^{2}}\underset{\varepsilon\rightarrow0^{+}}{\lim}\underset{-1+\frac{\varepsilon}{2}}{\overset{1-\frac{\varepsilon}{2}}{\int}}\frac{2e_{n}^{\left(\alpha\right)}\left(1+\varepsilon,x\right)}{1-x^{2}}dx=\left[\frac{1}{6}\left(\frac{1}{n}-n\right)-\frac{\alpha^{2}}{2\pi^{2}n}\right]\ln\mathcal{L}.
\end{equation}
Changing variables and taking the limit $\varepsilon\rightarrow0^{+}$
as before, the second part of the integral turns out to be
\begin{align}
 & -\frac{1}{\pi^{2}}\underset{\varepsilon\rightarrow0^{+}}{\lim}\underset{-1+\frac{\varepsilon}{2}}{\overset{1-\frac{\varepsilon}{2}}{\int}}\left[\psi\left(\frac{1}{2}+iW\left(x\right)\right)+\psi\left(\frac{1}{2}-iW\left(x\right)\right)\right]\frac{e_{n}^{\left(\alpha\right)}\left(1+\varepsilon,x\right)}{1-x^{2}}dx=\nonumber \\
 & =-\frac{i}{\pi}\underset{-\infty}{\overset{\infty}{\int}}\ln\left[\frac{2\cosh\left(\frac{nu}{2}+i\frac{\alpha}{2}\right)}{\left(2\cosh\left(\frac{u}{2}\right)\right)^{n}}\right]\frac{d}{du}\ln\frac{\Gamma\left(\frac{1}{2}+\frac{u}{2\pi i}\right)}{\Gamma\left(\frac{1}{2}-\frac{u}{2\pi i}\right)}du=\nonumber \\
 & =-\frac{1}{\pi^{2}}\underset{0}{\overset{\infty}{\int}}\ln\left[\frac{2\cos\alpha+2\cosh\left(nu\right)}{\left(2\cosh\left(\frac{u}{2}\right)\right)^{2n}}\right]du\underset{0}{\overset{\infty}{\int}}\left[\frac{e^{-t}}{t}-\frac{\cos\left(\frac{ut}{2\pi}\right)}{2\sinh\left(\frac{t}{2}\right)}\right]dt.
\end{align}
Finally, we arrive at
\[
\ln S_{n}^{\left(0\right)}\left(\alpha\right)=i\frac{k_{F}}{\pi}\alpha L+\left[\frac{1}{6}\left(\frac{1}{n}-n\right)-\frac{\alpha^{2}}{2\pi^{2}n}\right]\ln\mathcal{L}+\Upsilon_{0}\left(n,\alpha\right),
\]
where
\begin{equation}
\Upsilon_{0}\left(n,\alpha\right)\equiv-\frac{1}{\pi^{2}}\underset{0}{\overset{\infty}{\int}}\ln\left[\frac{2\cos\alpha+2\cosh\left(nu\right)}{\left(2\cosh\left(\frac{u}{2}\right)\right)^{2n}}\right]du\underset{0}{\overset{\infty}{\int}}\left[\frac{e^{-t}}{t}-\frac{\cos\left(\frac{ut}{2\pi}\right)}{2\sinh\left(\frac{t}{2}\right)}\right]dt.
\end{equation}

\subsection{Gaussian approximation of the charge distribution (XX model)\label{subsec: appendix 2}}

We write
\begin{equation}
\Upsilon_{0}\left(n,\alpha\right)=-\frac{1}{\pi^{2}}\underset{0}{\overset{\infty}{\int}}\ln\left[\frac{2\cos\alpha+2\cosh\left(nu\right)}{\left(2\cosh\left(\frac{u}{2}\right)\right)^{2n}}\right]du\underset{0}{\overset{\infty}{\int}}\left[\frac{e^{-t}}{t}-\frac{\cos\left(\frac{ut}{2\pi}\right)}{2\sinh\left(\frac{t}{2}\right)}\right]dt=c_{0}\left(n\right)+c_{2}\left(n\right)\alpha^{2}+\mathcal{O}\left(\alpha^{4}\right),
\end{equation}
and prove that $c_{2}\left(1\right)=-\frac{1+\gamma_{E}}{2\pi^{2}}$.
Indeed, substituting $n=1$,
\begin{equation}
\ln\left[\frac{2\cos\alpha+2\cosh\left(u\right)}{\left(2\cosh\left(\frac{u}{2}\right)\right)^{2}}\right]=\ln\left[\frac{2+2\cosh\left(u\right)}{\left(2\cosh\left(\frac{u}{2}\right)\right)^{2}}\right]-\frac{1}{4\cosh^{2}\left(\frac{u}{2}\right)}\alpha^{2}+\mathcal{O}\left(\alpha^{4}\right),
\end{equation}
and so
\begin{equation}
c_{2}\left(1\right)=\frac{1}{4\pi^{2}}\underset{0}{\overset{\infty}{\int}}\frac{1}{\cosh^{2}\left(\frac{u}{2}\right)}du\underset{0}{\overset{\infty}{\int}}\left[\frac{e^{-t}}{t}-\frac{\cos\left(\frac{ut}{2\pi}\right)}{2\sinh\left(\frac{t}{2}\right)}\right]dt.
\end{equation}
We use
\begin{equation}
\underset{0}{\overset{\infty}{\int}}\frac{\cos\left(\frac{ut}{2\pi}\right)}{\cosh^{2}\left(\frac{u}{2}\right)}du=\underset{-\infty}{\overset{\infty}{\int}}\frac{e^{i\frac{t}{\pi}x}}{\cosh^{2}\left(x\right)}dx,
\end{equation}
where the complex integral can be calculated using a rectangular contour
with infinite horizontal sides at $\Im z=0$ and $\Im z=i\pi$, so
that we get
\begin{equation}
\underset{0}{\overset{\infty}{\int}}\frac{\cos\left(\frac{ut}{2\pi}\right)}{\cosh^{2}\left(\frac{u}{2}\right)}du=\frac{2te^{-\frac{t}{2}}}{1-e^{-t}}.
\end{equation}
We can therefore write
\begin{align}
c_{2}\left(1\right) & =\frac{1}{2\pi^{2}}\underset{0}{\overset{\infty}{\int}}\left[\frac{e^{-t}}{t}-\frac{te^{-\frac{t}{2}}}{\left(1-e^{-t}\right)\left(e^{\frac{t}{2}}-e^{-\frac{t}{2}}\right)}\right]dt=\nonumber \\
 & =\frac{1}{2\pi^{2}}\underset{0}{\overset{\infty}{\int}}\left[\frac{1-e^{-t}-t}{t\left(e^{t}-1\right)}+\frac{1}{e^{t}-1}-\frac{te^{t}}{\left(e^{t}-1\right)^{2}}\right]dt=\nonumber \\
 & =\frac{-\gamma_{E}}{2\pi^{2}}+\frac{1}{2\pi^{2}}\underset{0}{\overset{\infty}{\int}}\frac{d}{dt}\left(\frac{t}{e^{t}-1}\right)dt=\nonumber \\
 & =-\frac{\gamma_{E}+1}{2\pi^{2}},\\
\nonumber 
\end{align}
where we have used the identity $\gamma_{E}=\underset{0}{\overset{\infty}{\int}}\frac{e^{-t}+t-1}{t\left(e^{t}-1\right)}dt$
\cite{whittaker_watson_1996}. 

\subsection{Decomposition of the vNEE (gapped XY model)\label{subsec: appendix 3}}

We present here a detailed calculation of $\mathcal{S}^{\left(-\right)}=-\underset{n\rightarrow1}{\lim}\partial_{n}S_{n}^{\left(-\right)}$
as $L\rightarrow\infty$, based on the result for $S_{n}^{\left(-\right)}$
in (\ref{eq: XY Renyi symmetry block}). For $h<2$ we obviously have
${\cal S}^{\left(-\right)}\rightarrow0$. For $h>2$, we can calculate
the derivative of the expression for $S_{n}^{\left(-\right)}$ by
rewriting it in terms of the Jacobi theta functions:
\begin{equation}
\left(-1\right)^{L}S_{n}^{\left(-\right)}\rightarrow\left[\frac{\left(kk'\right)^{2n}k_{n}'^{4}}{16^{n-1}k_{n}^{2}}\right]^{\frac{1}{12}}=\left[\frac{\theta_{2}^{4n}\left(q\right)\theta_{4}^{4n}\left(q\right)\theta_{4}^{8}\left(q^{n}\right)}{16^{n-1}\theta_{3}^{8n}\left(q\right)\theta_{2}^{4}\left(q^{n}\right)\theta_{3}^{4}\left(q^{n}\right)}\right]^{\frac{1}{12}}.
\end{equation}
 After some elementary steps, we arrive at 
\begin{equation}
\left(-1\right)^{L}\mathcal{S}^{\left(-\right)}\rightarrow\frac{\sqrt{k'}}{3}\left[\ln2-\frac{1}{2}\ln\left(k\cdot k'\right)+q\ln q\cdot\left(\frac{\theta_{3}'\left(q\right)}{\theta_{3}\left(q\right)}+\frac{\theta_{2}'\left(q\right)}{\theta_{2}\left(q\right)}-\frac{2\theta_{4}'\left(q\right)}{\theta_{4}\left(q\right)}\right)\right],
\end{equation}
where $\theta_{j}'\left(q\right)\equiv\frac{d}{dq}\theta_{j}\left(q\right)$.
For further simplification, we use the fact that
\begin{equation}
\frac{\theta_{3}'\left(q\right)}{\theta_{3}\left(q\right)}+\frac{\theta_{2}'\left(q\right)}{\theta_{2}\left(q\right)}-\frac{2\theta_{4}'\left(q\right)}{\theta_{4}\left(q\right)}=\frac{d}{dq}\ln\left(\frac{\theta_{2}\theta_{3}}{\theta_{4}^{2}}\right)=\frac{d}{dq}\ln\left(\frac{k^{\frac{1}{2}}}{k'}\right),
\end{equation}
along with the identity \cite{whittaker_watson_1996}
\begin{equation}
\frac{k^{\frac{1}{2}}}{k'}=2q^{\frac{1}{4}}\underset{m=1}{\overset{\infty}{\Pi}}\left(1+q^{m}\right)^{6},
\end{equation}
in order to obtain that 
\begin{equation}
q\left(\frac{\theta_{3}'\left(q\right)}{\theta_{3}\left(q\right)}+\frac{\theta_{2}'\left(q\right)}{\theta_{2}\left(q\right)}-\frac{2\theta_{4}'\left(q\right)}{\theta_{4}\left(q\right)}\right)=\frac{1}{4}+6\underset{m=1}{\overset{\infty}{\sum}}\frac{mq^{m}}{1+q^{m}}.
\end{equation}
To calculate the sum of the remaining series, we use \cite{2018arXiv180508648H}
\begin{equation}
\theta_{3}^{4}\left(q\right)=1+8\underset{m=1}{\overset{\infty}{\sum}}\frac{mq^{m}}{1+\left(-q\right)^{m}},
\end{equation}
and also $\theta_{4}\left(q\right)=\theta_{3}\left(-q\right)$ and
$\theta_{2}^{4}+\theta_{4}^{4}=\theta_{3}^{4}$, in order to arrive
at
\begin{align}
\frac{1}{24}\left(\theta_{3}^{4}+\theta_{2}^{4}-1\right) & =\frac{1}{24}\left(2\theta_{3}^{4}-\theta_{4}^{4}-1\right)=\nonumber \\
 & =\frac{1}{3}\underset{m=1}{\overset{\infty}{\sum}}\left[\frac{2mq^{m}}{1+\left(-q\right)^{m}}-\frac{m\left(-q\right)^{m}}{1+q^{m}}\right]=\nonumber \\
 & =\underset{m=1}{\overset{\infty}{\sum}}\frac{mq^{m}}{1+q^{m}}.
\end{align}
Additionally, we note that the following identity holds \cite{NIST:DLMF}:
\begin{equation}
I\left(k\right)=\frac{\pi}{2}\theta_{3}^{2}\left(q\right).
\end{equation}
We can therefore write
\begin{equation}
\frac{1}{4}+6\underset{m=1}{\overset{\infty}{\sum}}\frac{mq^{m}}{1+q^{m}}=\frac{1}{4}\left(\theta_{3}^{4}\left(q\right)+\theta_{2}^{4}\left(q\right)\right)=\frac{I^{2}\left(k\right)}{\pi^{2}}\left(1+k^{2}\right),
\end{equation}
and consequently we obtain for $h>2$ that
\begin{equation}
\left(-1\right)^{L}\mathcal{S}^{\left(-\right)}\rightarrow\frac{\sqrt{k'}}{3}\left[\ln2-\frac{1}{2}\ln\left(k\cdot k'\right)-\frac{I\left(k\right)I\left(k'\right)}{\pi}\left(1+k^{2}\right)\right].
\end{equation}

\bibliography{Manuscript}

\end{document}